\documentclass[aps,prb,twocolumn,superscriptaddress]{revtex4}
\usepackage{graphicx}
\usepackage{latexsym}
\usepackage{amssymb}
\usepackage{amsmath}
\usepackage{amsfonts}
\usepackage{upgreek}
\usepackage{bm}
\usepackage{multirow}
\usepackage{soul}
\usepackage{enumitem}
\usepackage{color}
\usepackage{youngtab}
\newcommand{\bra}[1]{\langle #1|}
\newcommand{\ket}[1]{|#1 \rangle}
\newcommand{\dd}{\mathrm{d}}
\newcommand{\ii}{\mathrm{i}}
\newcommand{\SU}{\mathrm{SU}}
\newcommand{\U}{\mathrm{U}}
\renewcommand{\O}{\mathrm{O}}
\newcommand{\SO}{\mathrm{SO}}
\newcommand{\Spin}{\mathrm{Spin}}
\newcommand{\Sp}{\mathrm{Sp}}
\newcommand{\dsZ}{\mathbb{Z}}
\newcommand{\dsR}{\mathbb{R}}

\newcommand{\Tr}{\operatorname{Tr}}

\renewcommand{\Re}{\operatorname{Re}}
\renewcommand{\Im}{\operatorname{Im}}
\newcommand{\vect}[1]{{\bm{#1}}}

\newcommand{\mat}[1]{\left[\begin{matrix}#1\end{matrix}\right]}
\newcommand{\eq}[1]{\begin{equation}#1\end{equation}}
\newcommand{\eqs}[1]{\begin{equation}\begin{split}#1\end{split}\end{equation}}
\newcommand{\eqnref}[1]{Eq.\,\eqref{#1}}
\newcommand{\figref}[1]{Fig.\,\ref{#1}}

\begin{document}

\title{Symmetric Fermion Mass Generation as Deconfined Quantum Criticality}
\author{Yi-Zhuang You}
\author{Yin-Chen He}
\affiliation{Department of Physics, Harvard University, Cambridge, MA 02138, USA}
\author{Cenke Xu}
\affiliation{Department of Physics, University of California,
Santa Barbara, CA 93106, USA}
\author{Ashvin Vishwanath}
\affiliation{Department of Physics, Harvard University, Cambridge, MA 02138, USA}

\date{\today}
\begin{abstract}
Massless 2+1D Dirac fermions arise in a variety of systems from graphene to the surfaces of topological insulators, where generating a mass is typically associated with breaking a symmetry. However, with strong interactions, a symmetric gapped phase can arise for multiples of eight Dirac fermions. A continuous quantum phase transition from the massless Dirac phase to this massive phase, which we term Symmetric Mass Generation (SMG), is necessarily beyond the Landau paradigm and is hard to describe even at the conceptual level. Nevertheless, such transition has been consistently observed in several numerical studies recently. Here, we propose a theory for the SMG transition which is reminiscent of deconfined criticality and involves emergent non-Abelian gauge fields coupled both to Dirac fermions and to critical Higgs bosons.  We motivate the theory using an explicit parton construction and discuss predictions for numerics. Additionally, we show that the fermion Green's function is expected to undergo a zero to pole transition across the critical point. 
\end{abstract}
\maketitle

\emph{Introduction---}
Recently, much attention has been lavished on band structures with symmetry protected nodal points (Dirac and Weyl semimetals)\cite{Wallace:1947ei,Divincenzo:1984xu,Murakami:2007ye,Wan:2011rp,Burkov:2011ez,Burkov:2011rp,Young:2012pl} in both two\cite{Novoselov:2004hi,Novoselov:2005ns,Zhang:2005kl} and three spatial dimensions.\cite{Borisenko:2014yb,Weng:2015ay,Huang:2015or,Lv:2015rv,Xu:2015eh} The paradigmatic example is graphene, where the band touching points are protected by symmetry, and the low energy dispersion around these points is captured by the massless 2D Dirac equation.\cite{Castro-Neto:2009dy} Similarly, massless Dirac fermions also appear on the surface of free fermion topological phases\cite{Franz:2013la}. A key question pertains to the stability of the Dirac nodes in the presence of interactions. This controls whether the materials remains a semimetal or develops a gap leading to a semiconductor. Typically, this has been discussed in terms of interaction induced symmetry lowering, where interactions lead to a spontaneous symmetry breaking. The resulting lowering of symmetry allows for an energy gap. The physics in these settings can be modeled by a mean field ``mass'' term that is spontaneously generated on lowering the symmetry, and gaps out the Dirac fermions. This is the standard mass generation in the Gross-Neveu\cite{Gross:1974fc} and the Yukawa-Higgs models. The main challenge then is identifying the appropriate channel of symmetry breaking, following which one can utilize the Landau paradigm of order parameters to describe the mass generation. 

In this work we will discuss an altogether different mechanism of mass generation for Dirac fermions, that breaks no symmetries and cannot be modeled by a single-particle mass term at the free fermion level. The possibility of such a scenario is informed by recent developments in the theory of interacting fermionic symmetry protected topological (SPT) phases,\cite{Fidkowski:2010bf,Fidkowski:2011dd,Ryu:2012ph,Qi:2013qe,Yao:2013yg,Fidkowski:2013ww,Wang:2014lm,Gu:2014tw,You:2014vp,Yoshida:2015aj,Gu:2015cy,Song:2016ut,Queiroz:2016se,Wang:2017ty} relating to the stability of free fermion topological insulators/superconductors to interactions. The paradigmatic example given by Fidkowski and Kitaev\cite{Fidkowski:2010bf,Fidkowski:2011dd} is the 1+1D Majorana chain with an appropriately defined time reversal that protects edge Majorana modes regardless of their multiplicity. However interactions lead to an energy gap to these modes when they are multiples of eight, leading to a reduction of the free fermion classification $\dsZ\rightarrow \dsZ_8$. More relevant to our purposes is the interaction reduction of 2+1D surface states of 3+1D topological phases, which contain Majorana or Dirac fermions. Indeed, here with the standard time reversal for electrons (class DIII),\cite{Fidkowski:2013ww,Wang:2014lm,Metlitski:2014fp} there is a interaction reduced classification $\dsZ\rightarrow \dsZ_{16}$ of topological superconductors, implying that sixteen surface Majorana fermions or equivalently eight Dirac fermions are unstable towards a massive (gapped) phase in the presence of strong interactions without breaking any symmetry. 

These considerations prompt us to look for a model of an electronic semimetal with eight Dirac nodes in 2D. A single layer of graphene with its two fold valley and two fold spin degeneracy leads to four Dirac nodes, hence we need to consider \emph{two layers}\footnote{To preserve the Dirac dispersion we avoid discussing Bernal stacked bilayer graphene. Instead we have in mind twisted sheets of graphene that reduce interlayer tunneling.} of graphene to obtain eight Dirac cones in all (by combining the valley, spin and layer degeneracies). There is a simple way to see that at half filling it is possible to realize a symmetric insulating phase if interactions are included. 
Let us consider an antiferromagnetic Heisenberg spin interaction $H_{\rm int} = J\sum_i \vect{S}_{i1}\cdot \vect{S}_{i2}$ between the vertically displaced sites across the two layers.\cite{Slagle:2015lo} Since on average we have one electron per site and each electron carries spin-1/2, the two electrons across the layer will pair into singlets and acquire an energy gap as long as the interlayer Heisenberg interaction is strong enough. This leads to a fully-gapped and non-degenerated ground state, which can be described as a direct product state of interlayer singlets. The state neither breaks any symmetry nor does it develop topological order.  Therefore it is a \emph{featureless gapped phase} in 2+1D.\cite{Kimchi:2013fk,Jiang:2015qv,Kim:2016mz} The strong-coupling interaction mass (the many-body gap) that the electrons acquire in this phase is called the \emph{symmetric mass},\cite{BenTov:2015lh} and the continuous phase transition (if it exists) between the Dirac semimetal and the featureless insulator will be called  \emph{symmetric mass generation} (SMG).\cite{He:2016qy,Catterall:2016sw,Ayyar:2016fi} Note, one can also discuss the transition for a system with fewer Dirac fermions. For the surface of a fermionic SPT phase (eg. in class DIII or AIII), the gapped phases necessarily involve topological order \cite{Fidkowski:2013ww,Wang:2014lm,Metlitski:2014fp} and constitute a rather different problem. For the intrinsically 2D system of graphene with an even number of sites in the unit cell, it is believed there is no intrinsic obstruction to realizing a trivial gapped phase (e.g.~as shown for spinful single layer graphene in Ref.\,\onlinecite{Kim:2016mz}). However writing Hamiltonian that realize these gapped phases is itself a nontrivial task. Therefore we focus on the case of 8 Dirac nodes in 2+1D systems where the gapped phases are readily accessible and the numerical evidence for a single continuous transition is encouraging. 

\begin{figure}[htbp]
\begin{center}
\includegraphics[width=0.42\textwidth]{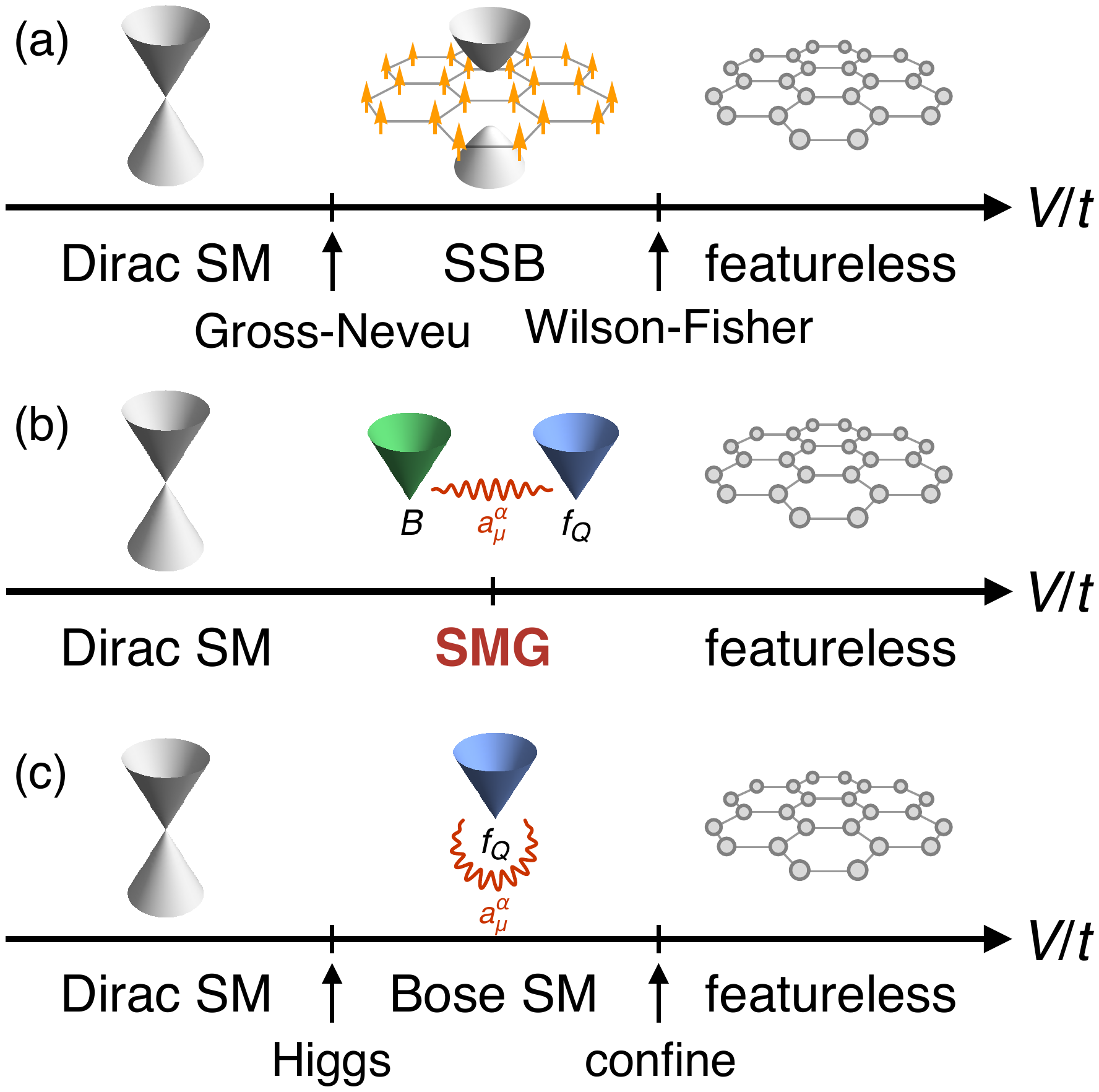}
\caption{Possible scenarios of transitions from the Dirac semimetal (Dirac SM) to featureless gapped phase. (a) Landau paradigm: an intermediate spontaneous symmetry breaking (SSB) phase sandwiched between the Gross-Neveu and the Wilson-Fisher transitions. (b) A direct continuous transition: the symmetric mass generation (SMG) as a deconfined quantum critical point, with emergent gauge field and fractionalized partons. (c) More exotic (and less likely) scenario: an intermediate Bose semimetal (Bose SM) critical phase between the Higgs and confinement transitions.}
\label{fig: scenarios}
\end{center}
\end{figure}

What are the possible scenarios for the transition from the Dirac semimetal to the featureless gapped phase? At least for small $J$, it is known that short-ranged interactions are perturbatively irrelevant for 2D Dirac fermions thus the transition can only occur at finite interaction strengths. Unlike the lower dimensional cases, where the instability of 1+1D gapless fermions is manifest perturbatively and/or can be studied with powerful tools such as bosonization, the situation for the 2+1D problem is more challenging. On general grounds, there could be several scenarios as we step out of the Dirac semimetal phase. First, there could simply be a direct first-order transition to the featureless gapped phase, where the symmetric mass gap opens up discontinuously. Next, an intervening symmetry breaking phase may occur, leading to an energy gap to the fermions. Subsequently the symmetry could be restored accomplishing the phase change in a two step process, as illustrated in \figref{fig: scenarios}(a). A different two step evolution involves the existence of an exotic critical phase that can be stable over a range of parameters, dubbed as the \emph{Bose semimetal} phase\cite{Lee:2005qq,Motrunich:2005fp,Sheng:2009hc,Block:2011ad} in \figref{fig: scenarios}(c). It is a gapless quantum liquid of bosons and can be described as a generalized Gutzwiller projected Dirac semimetal. The most interesting possibility is shown in \figref{fig: scenarios}(b), where the SMG occurs as a single continuous transition without any intermediate phases. Remarkably, numerical simulations of the problem in different models with various microscopic symmetries using different numerical methods\cite{Slagle:2015lo,He:2016qy,Catterall:2016sw,Ayyar:2016fi,Catterall:2016nh,Ayyar:2016tg,Ayyar:2016ph} seem to uniformly point towards a single continuous SMG transition. All these models share one key common property that the weakly interacting semimetal phase should have exactly \emph{eight} massless Dirac fermions. Even at the conceptual level it is unclear how to write down a theory for this putative transition. This is the problem addressed in this work.
 
Since the SMG transition lies outside the Landau symmetry breaking paradigm, it would necessarily be exotic and require new ideas. The strategy we will adopt is to consider a form of fractionalization, where the symmetry quantum number of the electron is peeled off from their Dirac dispersion and carried away by a set of bosonic partons, while the Dirac cone structure is still maintained by a set of symmetry neutral fermionic partons. The process of fractionalization leads to an emergent gauge interaction between the bosonic and fermionic partons. In this framework, the semimetal phase corresponds to the condensed (Higgs) phase of the bosons. The featureless gapped phase corresponds to the symmetric gapped phase  of the bosons, which triggers gauge confinement of the remaining degrees of freedoms. 
This theory of SMG therefore falls in the category of deconfined quantum critical points,\cite{Senthil:2004wj,Motrunich:2004hh,Senthil:2004qm} that contains a non-Abelian (Yang-Mills) gauge field coupled to both the massless scalar (Higgs) fields and eight flavors of massless Dirac fermions. 

In the following, we will first introduce a minimal model for the SMG in 2D with $\SU(4)$ symmetry.
We will develop an intuitive picture of the gapped phase as a paired superconductor in which fluctuations restore symmetry but leave the gap intact. This will motivate our parton construction and lead to a field theory description for SMG. Finally we discussed the implication of our theory for the fermion Green's function which can be tested in numerics.

\emph{Model---}
Consider the honeycomb lattice with four flavors of fermions at half filling on each site. This is a model of two layers of graphene (each with two component spinful fermions). Previously we discussed how an interlayer spin interaction could lead to singlets, but it will be useful to enhance the symmetry and consider the fermions to be fully symmetric under the $\SU(4)$ rotation of the four flavors. A minimal model that captures this is given by:
\begin{eqnarray} 
\label{eq: H0}
H &=& H_0+H_I \nonumber \\
H_0 &=& -t\sum_{\langle ij \rangle}\sum_{a=1}^4\big (c^\dagger_{ia}c_{ja} + \text{h.c.}\big)
\end{eqnarray}
where $c_{ia}$ is the fermion operator on site $i$ and of flavor $a=1,2,3,4$. Now consider the interaction term which preserves $\SU(4)$ symmetry:
\eq{\label{eq: Hint}
H_I =-V\sum_i \big( c^\dagger_{i1}c^\dagger_{i2}c^\dagger_{i3}c^\dagger_{i4} + \text{h.c.} \big).}
Note however the $V$ term does not preserve the charge conservation of the fermions. The charge $\U(1)$ symmetry is explicitly broken at the Hamiltonian level and excluded from our symmetry consideration. This may be interpreted as a proximity induced charge-`4e' superconductivity.\cite{Kivelson:1990wy,Berg:2009uq,Radzihovsky:2009bk,Berg:2009nm,Moon:2012ng,Jiang:2016km}. If the regular (charge-2e) superconductivity was brought in proximity to graphene, the fermions would immediately be gapped. In contrast  the presence of weak four fermion terms $V\ll t$ do not destabilize the Dirac cone and a finite interaction strength is needed for a transition to occur. On the other hand, in the strong coupling limit $V \gg t$, the ground state is a simple product state of on-site fluctuating charge-4e quartets:
\begin{equation}\label{eq: SC4e}
\ket{\Psi_c} = \prod_i \big (1+c^\dagger_{i1}c^\dagger_{i2}c^\dagger_{i3}c^\dagger_{i4} \big)\ket{0_c},
\end{equation}
where $\ket{0_c}$ denotes the fermion vacuum state. Therefore a  transition is expected between the gapless Dirac semimetal and the gapped charge-4e superconductor, as we tune the interaction strength. Numerical simulations of the $\SU(4)$ symmetric model\cite{He:2016qy,Catterall:2016sw,Ayyar:2016fi,Catterall:2016nh,Ayyar:2016tg,Ayyar:2016ph} point to a single continuous transition, i.e. the SMG transition. In the following, we will build a theory for it. One can also think that the charge-4e interaction $H_I$ is related to an interlayer spin-spin interaction as motivated in the introduction by suitable particle hole transformation of two of the four fermion components. The only modification is that we need to only consider the XY components of the interlayer spin interaction. There are several other choices of interactions\cite{Fidkowski:2010bf,Fidkowski:2011dd,Slagle:2015lo,You:2015lj} that also drive the SMG transition, but for this work, we will only focus on the charge-4e interaction $H_I$ described in \eqnref{eq: Hint}.  

{\em Symmetries---} Symmetries of the model include not only the $\SU(4)$ internal symmetry but also the lattice symmetry and the particle-hole symmetry which fixes half filling. The \emph{lattice symmetry} $G_\text{latt}$ includes translation, rotation and reflection symmetries of the honeycomb lattice. The particle-hole symmetry $\dsZ_2^\mathcal{S}$ acts as $\mathcal{S}:c_i\to(-)^ic_i^\dagger$ followed by complex conjugation, such that $\mathcal{S}^2=+1$, which is also known as the \emph{chiral symmetry} or the CT symmetry.\cite{Schnyder:2008os,Ryu:2010fe,Ludwig:2016pt}  The combined symmetry $G_\text{latt}\times\SU(4)\rtimes\dsZ_2^\mathcal{S}$ protects the Dirac semimetal from all fermion bilinear masses and the chemical potential shift. This can be seen from the field theory description for the Dirac semimetal
$\mathcal{L}=\sum_{Q=K,K'}\bar{c}_{Q}\gamma^\mu\ii\partial_\mu c_{Q}$,
where $c_Q$ is a $\SU(4)$ fundamental spinor at each valley ($Q=K,K'$). The $\SU(4)$ symmetric bilinear mass terms must take the form of $\bar{c}_QM_{QQ'}c_{Q'}$ with a Hermitian $2\times 2$ matrix $M$ in the valley sector. The space of $M$ is spanned by four Pauli matrices (including $\sigma^0$) basis, which correspond respectively to the Chern insulator gap, the charge density wave gap, and two Kekul\'e dimerization gaps. The first two break the particle-hole and the reflection symmetries and the last two break the translation symmetry, so none of them is allowed by the full symmetry. So the remaining option to generate fermion masses without breaking any symmetry is to invoke fermion interactions, such as the charge-4e interaction $H_I$.

\emph{Featureless Gapped Phase---}
To understand the SMG transition, we need to first understand both sides of the transition. The Dirac semimetal phase is relatively simple. As the interaction is weak and irrelevant, the semimetal phase is well described by the  fermion band theory. The featureless gapped phase (the charge-4e superconductor) is more exotic. As the gap is of the many-body nature, it can not be described by the simple band theory picture. Nevertheless, much understanding of the charge-4e superconductor was obtained by disordering the charge-2e superconductor in previous studies.\cite{Kivelson:1990wy,Berg:2009uq,Radzihovsky:2009bk,Berg:2009nm} We will take the same approach here. Let us consider fermion mass generation in two steps: we first gap the fermion by introducing the charge-2e pairing at the price of breaking the  symmetry, and then we restore the symmetry by disordering the pairing field. The discussion will lead to a parton construction for the featureless gapped phase, based on which we can further explore the possibility to merge the two steps of the mass generation into one single transition \emph{without} the intermediate symmetry breaking phase.

Let us start from the semimetal side and consider the $\SU(4)$ sextet pairing on each site (which has six components labeled by $m=1,\cdots,6$) \cite{Wu:2006gf,You:2015lj}
\eq{\label{eq: Delta}\Delta_i^m=\frac{1}{2}\sum_{a,b}c_{ia} \beta^{m}_{ab}c_{ib},}
where $\beta^m$ are antisymmetric $4\times4$ matrices given by
$\vect{\beta}=(\sigma^{12},\sigma^{20},\sigma^{32},\ii\sigma^{21},\ii\sigma^{02},\ii\sigma^{23})$,
where $\sigma^{\mu\nu}=\sigma^\mu\otimes\sigma^\nu$ denotes the direct product of Pauli matrices $\sigma^\mu$ and $\sigma^\nu$. The paring operator $\vect{\Delta}_i$ rotates like an $\O(6)$ vector under the $\SU(4)\cong\Spin(6)$ symmetry, and it transforms under the chiral symmetry as $\mathcal{S}: \vect{\Delta}_i\to-\vect{\Delta}_i^\dagger$. Introducing such a flavor sextet pairing to the Hamiltonian
\eq{H_{\vect{M}}=-\sum_{i}\vect{M}\cdot(\vect{\Delta}_i+\vect{\Delta}_i^\dagger)} will break the $\SU(4)$ symmetry (down to its $\Sp(2)\cong\Spin(5)$ subgroup) as well as the chiral symmetry $\dsZ_2^\mathcal{S}$, and at the same time gap out all the Dirac fermions. In the limit that the paring gap $|\vect{M}|\to\infty$, the fermion correlation length shrinks to zero, and the ground state wave function (of $H_{\vect{M}}$) reads
\eq{\label{eq: Psi_cM}\ket{\Psi_{c,\vect{M}}}=\prod_{i}\big(1+\hat{\vect{M}}\cdot\vect{\Delta}_i^\dagger +c_{i1}^\dagger c_{i2}^\dagger c_{i3}^\dagger c_{i4}^\dagger\big)\ket{0_c},}
where $\hat{\vect{M}}=\vect{M}/|\vect{M}|$ is the unit vector that points out the ``direction'' of the sextet pairing. Comparing $\ket{\Psi_{c,\vect{M}}}$ with the wave function $\ket{\Psi_c}$ for the featureless gapped phase in \eqnref{eq: SC4e}, we can see that the most essential difference lies in the additional fermion bilinear term $\hat{\vect{M}}\cdot \vect{\Delta}_i^\dagger$ in $\ket{\Psi_{c,\vect{M}}}$, which breaks the $\SU(4)$ symmetry.

To restore the $\SU(4)$ symmetry, we need to remove the fermion bilinear term from the wave function. This amounts to symmetrizing the wave function $\ket{\Psi_{c,\vect{M}}}$ over all directions of $\vect{M}$, or in other words, projecting the wave function $\ket{\Psi_{c,\vect{M}}}$ to the $\SU(4)$ symmetric subspace. Loosely speaking, we propose the following projective construction
\eq{\label{eq: projective}\ket{\Psi_{c}}\sim\int_{S^5}\dd\vect{M}\ket{\Psi_{c,\vect{M}}}.} 
This construction will be made precise using the parton formalism shortly.  But the lesson we learnt is that the fermion bilinear mass $\vect{M}$ serves as a convenient scaffold to construct the featureless gapped state, which can be removed by the symmetrization in the end.

\emph{Parton Construction---}
The idea of carrying out the symmetrization on every site invites us to think about ``gauging'' the $\SU(4)$ symmetry as follows. 
Consider decomposing the physical fermion $c_{ia}$ into bosonic $B_{iab}$ and fermionic $f_{ib}$ partons ($a,b=1,2,3,4$)
\eq{\label{eq: c=bf} c_{ia}=\sum_{b=1}^4B_{iab}f_{ib},}
with the ``orthogonal constraint''\footnote{To preserve the fermion anticommutation relation, we may expect the bosonic parton $B_{i}$ to form an ``unitary matrix'': $\sum_{c}B_{ica}^\dagger B_{icb}=\sum_{c}B_{iac}B_{ibc}^\dagger=\delta_{ab}$. However this constraint turns out to be too strong, which restrict the boson Hilbert space to one dimension and effectively quench all the bosonic degrees of freedom, i.e. there is only one state $\det(B_{i}^\dagger)\ket{0_B}$ that satisfies the constraint. So we choose to relax the constraint by allowing the boson number to fluctuate, meaning that the we still keep the orthogonality but abandon the normalization of the $B_{i}$ matrix. With this, the physical fermion and the fermionic parton are still related by a unitary transform but up to a constant. The constant may be interpreted as the quasiparticle weight.} on the bosonic parton Hilbert space
$\forall a\neq b:\sum_{c}B_{ica}^\dagger B_{icb}=\sum_{c}B_{iac}B_{ibc}^\dagger=0$.
An $\SU(4)$ gauge freedom emerges from the fractionalization.\footnote{Naively one may expect a larger $\U(4)$ gauge structure just by looking at the fractionalization scheme, but as we will see soon, the interaction terms of the bosonic partons can break the gauge group down to $\SU(4)$. In a sense, we choose to fix the $\U(1)$ gauge.} On each site, the  gauge transformation $U_{i}\in\SU(4)$ is implemented as
\eq{B_{iab}\to \sum_{c=1}^4B_{iac}U_{ibc}^*,\quad f_{ia}\to \sum_{b=1}^4 U_{iab}f_{ib}.}
Both the bosonic and the fermionic partons carry the $\SU(4)$ gauge charge. Besides the gauge charge, the $\SU(4)$ symmetry charge is carried solely by the bosonic parton. The chiral symmetry $\dsZ_2^\mathcal{S}$ acts projectively on the partons as $\mathcal{S}:B_i\to -\ii B_i^\dagger, f_i\to \ii (-)^i f_i^\dagger$, where the factor $\pm\ii$ should be understood as the gauge transform in the $\dsZ_4$ center of the $\SU(4)$ gauge group. So we have $\mathcal{S}^2=-1$ for both bosonic and fermionic partons, in contrast to $\mathcal{S}^2=+1$ for the physical fermion. As we will show later, such a projective $\dsZ_2^\mathcal{S}$ action is required by the non-trivial projective symmetry group (PSG) of the parton mean field theory.

\emph{Wavefunction from Partons---} Motivated by the previous projective construction, we put the fermionic parton in a $\SU(4)$ gauge sextet superconducting state\footnote{Note that the parton  ground state $\ket{\Psi_f}$ is not gauge invariant as it depends on the gauge sextet $\vect{M}$. This is legitimate because $\ket{\Psi_f}$ is \emph{not} a physical state and $\vect{M}$ is \emph{not} a variational parameter to appear in the final construction of the physical ground state after the gauge projection.} in analogy to \eqnref{eq: Psi_cM}, 
\eq{\label{eq: Psi_f}\ket{\Psi_f}=\prod_{i}\big(1+\hat{\vect{M}}\cdot \vect{\Delta}_{i}^{\dagger}[f] +f_{i1}^\dagger f_{i2}^\dagger f_{i3}^\dagger f_{i4}^\dagger\big)\ket{0_f},}
where $\ket{0_f}$ denotes the fermionic parton vacuum state and $\vect{\Delta}_i[f]= \tfrac{1}{2}\sum_{a,b} f_{ia} \vect{\beta}_{ab}f_{ib}$ is the gauge sextet paring operator of the fermionic parton $f_{ia}$, which is similar to the flavor sextet pairing of the physical fermion in \eqnref{eq: Delta}. With this gauge sextet pairing, the $\SU(4)\cong\Spin(6)$ gauge group is broken down to its $\Sp(2)\cong\Spin(5)$ subgroup. However the $\SU(4)$ symmetry remains untouched, because the symmetry charge is now carried by the bosonic parton. More importantly, the parton state $\ket{\Psi_f}$ is also symmetric under the chiral symmetry $\dsZ_2^\mathcal{S}$ in the PSG sense,\footnote{Under projective action of $\dsZ_2^\mathcal{S}$, the fermionic parton is particle-hole conjugated followed by the $\dsZ_4$ gauge transformation $f_i^\dagger\to -\ii(-)^if_i$, and the parton vacuum state is sent to the fully occupied state $\ket{0_f}\to \prod_i f_{i1}^\dagger f_{i2}^\dagger f_{i3}^\dagger f_{i4}^\dagger\ket{0_f}$.} which is in contrast to the physical fermion state $\ket{\Psi_{c,\vect{M}}}$ in \eqnref{eq: Psi_cM} where $\dsZ_2^\mathcal{S}$ is broken. Using the previously proposed PSG transformation $\mathcal{S}:f_i\to \ii (-)^i f_i^\dagger$, it can be shown that the parton pairing operator transforms as $\mathcal{S}:\vect{\Delta}_i[f]\leftrightarrow \vect{\Delta}_i^\dagger [f]$. Hence the gauge sextet pairing term $\vect{M}\cdot(\vect{\Delta}_i[f]+ \vect{\Delta}_i^\dagger [f])$ is $\dsZ_2^\mathcal{S}$ symmetric, so as the resulting mean-field state in \eqnref{eq: Psi_f}. To construct a $\SU(4)$ symmetric state, we consider putting the bosonic parton in a short-range correlated $\SU(4)$ singlet state. In the extreme limit of zero correlation length, an $\SU(4)$ symmetric many-body state takes the following form:
\eq{\label{eq: Psi_B}\ket{\Psi_B}=\prod_{i}\Big(1 + \frac{1}{4!} \epsilon_{abcd}B_{ia1}^\dagger B_{ib2}^\dagger B_{ic3}^\dagger B_{id4}^\dagger\Big)\ket{0_B},}
where $\ket{0_B}$ denotes the bosonic parton vacuum state. $\epsilon_{abcd}$ is the antisymmetric (Levi-Civita) tensor of four indices, such that the flavor indices are antisymmetrized to form the $\SU(4)$ singlet.

Now we take both the bosonic and the fermionic parton wave functions and project them to the physical fermion Hilbert space, 
\eq{\label{eq: Psi_c proj}\ket{\Psi_c}=\mathcal{P}\ket{\Psi_B \Psi_f},}
where the projection operator maps each parton Fock state to the corresponding Fock state of physical fermions
\eq{\mathcal{P}=\prod_{i,a}\left(\ket{0_c}\bra{0_f 0_B}+c_{ia}^\dagger \ket{0_c}\bra{0_f 0_B}\sum_{b=1}^4B_{iab}f_{ib}\right).}
The resulting state $\ket{\Psi_c}$ in \eqnref{eq: Psi_c proj} is precisely the featureless gapped state in \eqnref{eq: SC4e}. This parton construction provides us one plausible picture of the featureless gapped phase: the fermionic parton is in a gauge sextet paired state, while the bosonic parton is in a $\SU(4)$  symmetric gapped state, and the remaining gauge degrees of freedom are confined. On the other hand, the Dirac semimetal phase also admits a simple picture in the parton formalism: if we put the fermionic parton in the same Dirac band structure as the physical fermion and condense the bosonic parton to the state $\langle B_{iab}\rangle=Z\delta_{ab}$ (with $Z$ acting like the quasi-particle weight), then the physical fermion will be identified to the fermionic parton  $c_{ia}=Z f_{ia}$ and retrieve the Dirac band structure. We will implement these insights in a field theory below.

\emph{Field Theory---}
What we learned from the parton construction is that the Dirac semimetal and the symmetric massive phase correspond respectively to the Higgs and the confined phases of an $\SU(4)$ gauge theory. Thus if there is a direct continuous transition between them, it is conceivable that the transition should be a deconfined critical point,\cite{Senthil:2004wj,Motrunich:2004hh,Senthil:2004qm} i.e.~the gauge theory will be deconfined at and only at the transition point. Therefore we propose the following field theory description for the symmetric mass generation,
\eqs{\label{eq: L}\mathcal{L}=&\mathcal{L}_f+\mathcal{L}_B,\\
\mathcal{L}_f=&\sum_{Q=K,K'}\bar{f}_Q\gamma^\mu(\ii\partial_{\mu}-a_\mu^m\tau^m)f_Q+\mathcal{L}_\text{int},\\
\mathcal{L}_B=&-\Tr B(\ii\partial_\mu-a_\mu^m\tau^m)^2B^\dagger+r\Tr B B^\dagger\\
&+u_1(\Tr B B^\dagger)^2+u_2\Tr(B B^\dagger)^2\\
&+u_3(\det B+\text{h.c.})+\cdots,}
which contains the matter fields of bosonic partons $B$ and fermionic partons $f_Q$ as well as the $\SU(4)$ gauge field $a_\mu^m\tau^m$. The matrices $\tau^m$  ($m=1,\cdots,15$) are $\SU(4)$ generators (as $4\times4$ Hermitian traceless matrices), and $(\gamma^0,\gamma^1,\gamma^2)=(\sigma^2,\sigma^1,\sigma^3)$. $\mathcal{L}_\text{int}$ contains short-range interactions of the fermionic parton which will be specified later in \eqnref{eq: L_int}. This interaction term is treated perturbatively, but it will play an important role to deform the fermionic sector from a pure quantum chromodynamics (QCD) theory, giving rise to possible instabilities of spontaneous mass generation for the fermionic parton in the symmetric gapped phase (as to be analyzed soon).  Furthermore the emergent $\U(1)$ symmetry corresponding to rotating the overall phase of the fermionic parton ($f_Q\to e^{\ii \theta}f_Q$) will also be broken by the interaction $\mathcal{L}_\text{int}$. 

To reformulate the fractionalization scheme \eqnref{eq: c=bf} at the field theory level, we start with the low-energy physical fermions $c_Q$ ($Q=K,K'$) around $K$ and $K'$ points of the Brillouin zone. Both of them transform under the  $\SU(4)$ symmetry as fundamental representations. We can fractionalize the physical fermion field to the parton fields as
$c_{Q}=B\cdot f_{Q}$,
where $f_{Q}$ is a four component fermion field (transforming as a $\SU(4)$ gauge fundamental) for each valley $Q$, and $B$ is a $4\times4$ matrix field that transforms under both the $\SU(4)$ symmetry (from left) and the $\SU(4)$ gauge symmetry (from right). Based on the fractionalization scheme of \eqnref{eq: c=bf}, we expect the matrix field $B$ to be unitary (up to normalization constant $Z$) on the lattice scale. The constraint may be imposed by a Lagrangian multiplier $\lambda \Tr (B B^\dagger -Z^2)^2$, which, under renormalization, leads to an effective potential for $B$ in the field theory, whose leading terms ($r$ and $u_{1,2}$ terms of $\mathcal{L}_B$) are given in \eqnref{eq: L}. The $u_3$ term is another $\SU(4)$ symmetric four-boson interaction, which explicitly breaks the $\U(1)$ symmetry of $B$ and can be viewed as a descendant of the charge-4e superconducting interaction $H_I$ in \eqnref{eq: Hint}.

{\em Symmetric Gapped Phase:} In the field theory \eqnref{eq: L}, the SMG transition is driven by $r$. When $r>0$, the bosonic parton is gapped, leaving the fermionic parton coupled to the $\SU(4)$ gauge field below the scale of the bosonic parton gap $\Delta_B$, described by the $N_f=2$ $\SU(4)$ QCD theory. We assume that this $\SU(4)$ QCD theory is confining.\cite{Grover:2012rw} The resulting confined phase will depend on additional details. For example, if we considered a pure SU(4) QCD, with no additional four fermion interactions, a $\U(1)$ symmetry of $f_Q\to e^{\ii \theta}f_Q$ (the baryon number conservation) will be present, and will not be broken in the confined phase, by the Vafa-Witten theorem.\cite{VafaWitten} Instead, chiral symmetry breaking is likely to occur, breaking the $\SU(2)$ valley symmetry. However, for our purposes it will be crucial to include the following four fermion interaction term in the form of the pair-pair interaction of the gauge sextet pairing $\vect{\Delta}[f]=f_{K}^\intercal\ii\gamma^0\vect{\beta}f_{K'}$,
\eq{\label{eq: L_int}\mathcal{L}_\text{int}=\frac{g}{2}\big(\vect{\Delta}[f]\cdot\vect{\Delta}[f]+\text{h.c.}\big).}
This interaction can be written as $f_1f_2f_3f_4 + {\rm h.c.}$ equivalently (which is $\SU(4)$ gauge neutral the same as $\vect{\Delta}\cdot\vect{\Delta}$), reminiscent of the charge-4e interaction $V$ between electrons in \eqnref{eq: Hint} that drives the transition.  Now, the $\U(1)$ baryon number is no longer a global symmetry of the theory and Vafa-Witten does not forbid mass generation in the $\U(1)$ breaking (gauge sextet pairing) channel.

If the $\SU(4)$ gauge fluctuation were absent, the short-range interaction $\mathcal{L}_\text{int}$ would be perturbatively irrelevant, given the negative engineering dimension $[g]=-1<0$ of the coupling $g$. As the $\SU(4)$ gauge fluctuation is included in the QCD theory, the scaling dimension $[g]$ can receive anomalous dimension corrections. By a controlled renormalization group (RG) analysis based on the $1/N_f$ expansion, detailed in Appendix~\ref{sec: RG}, we compute the scaling dimension $[g]=-1+80/(\pi^2N_f)$ to the $1/N_f$ order, implying that the interaction ${\mathcal{L}}_\text{int}$ could become relevant (i.e.\,$[g]>0$) as we push $N_f$ to $N_f=2$. As the interaction flows strong under RG, it will drive the condensation of the gauge sextet pairing and lead to a mass term $\vect{M}\cdot(\vect{\Delta}[f]+\text{h.c.})$ which will gap the fermionic partons, at the scale of $\Delta_f\sim|\vect{M}|$, and break the gauge group down to $\Sp(2)$. In the absence of matter field fluctuations below the energy scale $\Delta_f$, the non-Abelian $\Sp(2)$ gauge field will confine itself (at a confinement  scale $\Delta_a$). However the $\SU(4)$ symmetry remains unbroken, since the bosonic parton is gapped and disordered. The particle-hole symmetry $\mathcal{S}$ also remains unbroken because it is realized on the fermionic partons projectively, $\mathcal{S}:f_{Q}\to \ii f_{Q}^\dagger$, where the $\dsZ_4$ gauge transformation is crucial to undo the sign change of $\vect{M}$ originally caused by the particle-hole transformation. Thus the system is in the featureless gapped phase that preserves all symmetries. 

{\em Massless Dirac Phase:} When $r<0$, the bosonic parton condenses $\langle B\rangle\neq0$. A positive $u_2>0$ term would favor the condensate configuration $\langle B\rangle$ to be a unitary matrix (up to an overall factor $Z$). Thus we can alway choose $\langle B_{ab}\rangle=Z\delta_{ab}$ by $\SU(4)$ gauge transformations. This will identify the $\SU(4)$ symmetry with the $\SU(4)$ gauge group and Higgs out all gauge fluctuations. The system is then in the Dirac semimetal phase, described by $\mathcal{L}=\sum_{Q=K,K'}\bar{c}_{Q}\gamma^\mu\ii\partial_\mu c_{Q}$, where $c_Q=Z f_Q$ and $Z$ may be interpreted as the quasiparticle weight. Any short-range interaction $\mathcal{L}_\text{int}$ among the fermionic parton in \eqnref{eq: L} will become perturbatively irrelevant once the $\SU(4)$ gauge fluctuation is Higgs out by the condensation of the bosonic parton, and the corresponding interaction-driven instability (such as the gauge sextet pairing instability) will cease to exist. In this way, the RG relevance of the fermionic parton interaction $\mathcal{L}_\text{int}$ (and the fermionic parton mass generation) is controlled by the mass $r$ of the bosonic parton, so the SMG does not need fine tuning (other than the only driving parameter $r$).

{\em Critical Point:} At $r=0$, {\em both}  fermionic ($f_Q$) and bosonic ($B$) partons are gapless. Together, they screen the $\SU(4)$ gauge field more efficiently, hence reducing the tendency to confinement. This opens up the possibility a stable deconfined $\SU(4)$ QCD-Higgs theory, which could describe the SGM critical point. We note that the similar behavior, namely the gauge confinement being irrelevant\cite{Senthil:2004qm} at the critical point due to the gapless bosons, was also discussed in the deconfined phase transition\cite{Senthil:2004wj,Motrunich:2004hh} between the N\'eel and the valence bond solid phases. On moving away from the critical point into the phase where the Higgs fields are gapped, confinement takes over. 

\begin{figure}[htbp]
\begin{center}
\includegraphics[width=0.33\textwidth]{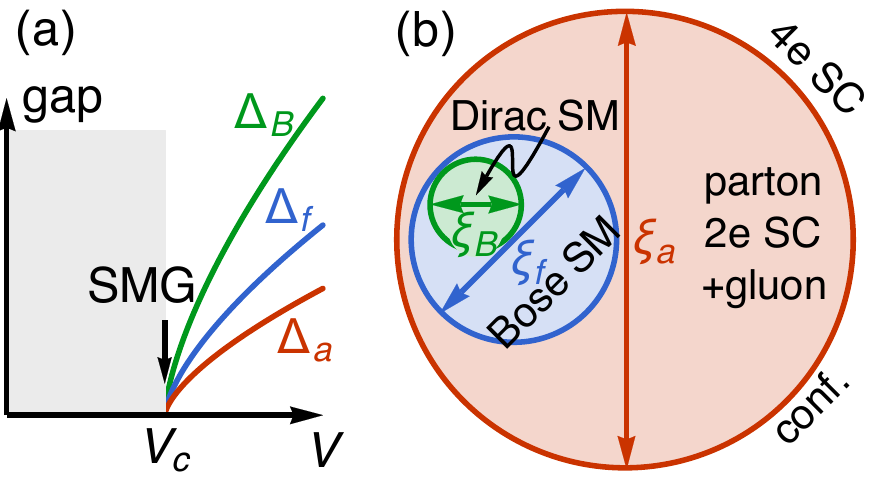}
\caption{(a) Catching-up energy scales of the bosonic parton gap $\Delta_B$, the fermionic parton gap $\Delta_f$, and the gauge gluon gap $\Delta_a$ on the confinement side of the SMG. (b) Hierarchical length scales of  $\xi_B$, $\xi_f$ and $\xi_a$ near the SMG transition.}
\label{fig: scales}
\end{center}
\end{figure}

Based on the above understanding, the energy scales $\Delta_B$, $\Delta_f$ and $\Delta_a$ should catch up one after another as we enter the featureless gapped phase, as illustrated in \figref{fig: scales}(a). This implies a hierarchy of length scales $\xi_{B,f,a}\sim\Delta_{B,f,a}^{-1}$ near the SMG transition from the side of the featureless gapped phase as shown in \figref{fig: scales}(b). For example, the $\SU(4)$ multiplet fluctuation will be gapped with the bosonic parton at the length scale of $\xi_B$. However the $\SU(4)$ singlet fluctuation can persist to a longer length scale $\xi_a$ until the $\Sp(2)$ gluon gets confined.

The above field theory description is in parallel with the parton construction as discussed previously. But the field theory also provides other possible scenarios for the transition(s) between the Dirac semimetal and the symmetric gapped phase, when the gaps $\Delta_B$, $\Delta_f$, $\Delta_a$ fail to open up together as the interaction $V/t$ increases. For example, by tuning the short-range interactions of the fermionic parton $f_Q$ in \eqnref{eq: L}, it is possible that the fermionic parton may develop the bilinear mass {\em before} the bosonic parton is gapped, an intermediate $\SU(4)$ symmetry breaking charge-2e  superconducting phase will set in, with the condensation of the $\SU(4)$ sextet Cooper pairs of the physical fermion, as shown in \figref{fig: scenarios}(a). Such a charge-2e superconducting phase was also observed in numerical simulations if the lattice model in Eq.\,(\ref{eq: H0},\ref{eq: Hint}) is deformed by the attractive Hubbard interaction\cite{Cai:2013ak,Cai:2013mq,Wang:2014kl} or by doping the chemical potential away from the Dirac point\cite{Jiang:2016km}. 

A more exotic scenario occurs if an extended deconfined phase is present, leading to an intermediate gapless quantum liquid. That is, if the fermion parton mass generation and the gauge confinement happens after the gapping of the bosonic parton,  as shown in \figref{fig: scenarios}(c). In this phase, the physical fermions are gapped, and the low-energy bosonic fluctuations are described by a wave function obtained from the gauge projection of the fermionic parton semimetal state. Therefore we may called it a \emph{Bose semimetal}  (BSM) phase.\cite{Lee:2005qq,Motrunich:2005fp,Sheng:2009hc,Block:2011ad} The gapless bosonic fluctuation should be $\SU(4)$ singlets and transform only under the lattice symmetry. A possible candidate is the valence bond solid order fluctuation. Dynamically, which of these scenarios are more favorable should depend on the details of parton interactions and gauge dynamics. Numerical evidence from the lattice model seems to support a direct continuous transition without either of the intermediate phases, as shown in \figref{fig: scenarios}(b). 

\emph{Fermion Green's Function---}
One smoking gun ``feature'' of the featureless gapped phase is the existence of zeroes in the fermion Green's function at zero frequency.\cite{You:2014br} To be precise, let us define the Green's function of the physical fermion to be $G_{ab}(x)=-\langle c_{a}(x)\bar{c}_{b}(0)\rangle$ where $x=(t,\vect{x})$ is the spacetime coordinate. Fourier transform to the momentum-frequency space, we have $G(k)=\int \dd^3x G(x)e^{\ii k_\mu x^\mu}$ with $k=(\omega,\vect{k})$. In the Dirac semimetal phase, poles of the Green's function appear along the band dispersion. In particular, at the $K$ and $K'$ points of the Brillouin zone where the fermion becomes gapless, the pole is pushed to zero frequency, and hence $G(k)\sim\omega^{-1}$. However in the symmetric gapped phase, as proven in Ref.\,\onlinecite{You:2014br}, the poles will be replaced by zeroes: $G(k)\sim \omega$ as $\omega\to0$ at $\vect{k}=K,K'$. In fact the Green's function zeros are symmetry protected in the featureless gapped phase, which was proved in Ref.\,\onlinecite{You:2014br}. Let us focus in the vicinity of the $K$ and $K'$ points and redefine $\vect{k}$ to be the momentum deviation from them. The SMG transition is also a zero-pole transition in the fermion Green's function at $k\to0$. Let us see how this is reproduced by the parton theory. 

\begin{figure}[htbp]
\begin{center}
\includegraphics[width=0.4\textwidth]{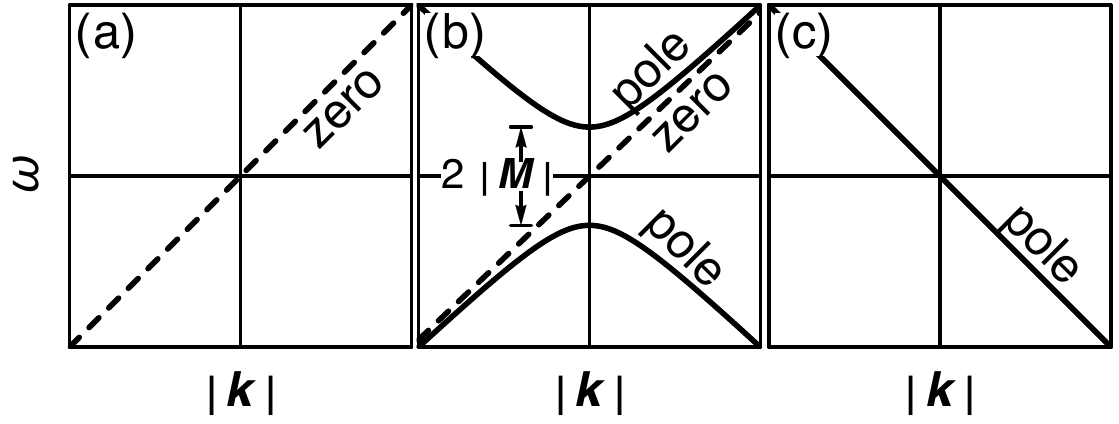}
\caption{The zero-pole transition of the fermion Green's function. Take the one of the $G(k)$ eigenvalues $(\omega-|\vect{k}|)/(\omega^2-|\vect{k}|^2-|\vect{M}|^2)$ for example.}
\label{fig: zero-pole}
\end{center}
\end{figure}

Using the parton construction outlined in \eqnref{eq: projective}, we can calculate the fermion Green's function deep in the featureless gapped phase. The result is (see Appendix~\ref{sec: G(k)} for details)
\eq{\label{eq: G(k)}G_{ab}(k)=\frac{\gamma^\mu k_\mu}{k^\mu k_\mu+\vect{M}^2}\delta_{ab}.}
where $|\vect{M}|$ corresponds to the sextet pairing gap of the fermionic parton.  In the featureless gapped phase (where $|\vect{M}|$ is finite), $G(k)$ approaches zero analytically at $k\to0$ as expected, see \figref{fig: zero-pole}(a). This lends confidence to our parton construction of the featureless gapped phase. The fact that the quasiparticle weight approaches to unity deep in the featureless gapped phase is also consistent with expectation. Because the charge-4e superconducting ground state is a fully gapped symmetric short-range entangled state (similar to a vacuum state), a physical fermion $c$ doped into the system will just propagate as a quasiparticle above its spectral gap set by the mass scale $|\vect{M}|$ without any fractionalization (which is also consistent with the picture that the $\SU(4)$ gauge theory is confining in the featureless gapped phase).

The Green's function $G(k)$ in \eqnref{eq: G(k)} also provides a plausible scenario for the zero-pole transition. As the gap $|\vect{M}|$ decreases, two branches of poles are brought down from high energy, as shown in \figref{fig: zero-pole}(b). They approach the line of zero asymptotically and eventually annihilate with the zero at the SMG transition where $|\vect{M}|\to0$. Then only a line of pole is left in the Dirac semimetal phase in \figref{fig: zero-pole}(c). A similar mechanism for the zero-pole transition was proposed in Ref.\,\onlinecite{Gurarie:2011gl}. However we also note that the Green's function $G(k)$ in \eqnref{eq: G(k)} can not describe the fermion correlation close to the SMG critical point, where the bosonic parton and the gauge fluctuation also become important, such that one needs to go beyond the variational approach to describe the vanishing quasiparticle weight and the continuum spectral function as a result of the fermion fractionalization.


\emph{Conclusion and Discussion---}
The SMG is an exotic quantum phase transition between the Dirac semimetal phase and a symmetric gapped phase, which cannot be understood within conventional theories of Dirac mass generation such as the Gross-Neveau mechanism. However, it has been numerically observed in different models with different interaction and symmetries. We propose a theoretical framework for SMG broadly as a deconfined quantum critical point, where the physical fermion is fractionalized into bosonic and fermionic partons with emergent gauge interaction. The Dirac semimetal phase corresponds to the Higgs phase and the featureless gapped phase corresponds to the confined phase. The gauge group, the parton flavor number and the symmetry assignment are flexible components of the theoretical framework that can be adapt to the model details, including the interaction parameters and the model symmetries. More work is required to understand what determines these parameters. 

In this work, we propose that  SMG in 2+1D with $\SU(4)$ global symmetry can be described by an $\SU(4)$ QCD$_3$-Higgs theory. Analyzing the nonperturbative dynamics of such a strongly coupled critical point is currently beyond our analytical capabilities, but a few statements can be made based on the basic structure of the theory following the approach used in $\U(1)$ deconfined quantum criticality\cite{Senthil:2006cs,Wang:2017zp}. At the critical point we expect an emergent SU(2) symmetry in the valley space, relating a pair of valence bond solid orders and staggered A/B sublattice order on the honeycomb lattice.  Potentially, there is an additional charge U(1) symmetry that could emerge at the critical point if either the $u_3\Re\det(B)$ in the bosonic parton sector or $\mathcal{L}_\text{int}$ of the fermionic partons is irrelevant at the transition. However this appears unlikely since the term that drives the transition, the four electron charge 4e superconductor term, itself breaks this symmetry. It will be interesting to test the enlarged symmetry of the SMG critical point in numerics. 
Another prediction is that the anomalous dimension of the electrons is large at criticality since it decays into a pair of partons.
Similarly, the $\SU(4)$ symmetry order parameters which are bilinears in the $B$ field  should also have large anomalous dimension. For example, the quantum Monte Carlo simulation in Ref.\,\onlinecite{He:2016qy} obtained the anomalous dimension $\eta_\text{SMG} = 0.7 \pm 0.1$ for the $\O(6)$ order parameter, much larger than $\eta_\text{WF}=0.035$ at the $\O(6)$ Wilson-Fisher fixed point.

The theoretical framework proposed in this work may be applied to the SMG with other symmetry groups and in other dimensions. For example, in a upcoming work,\cite{You:2017sy} we will study SMG in a model with lower symmetry,  $\SU(2)\times\SU(2)\times\SU(2)$ which will be described by a $\SU(2)$ QCD$_3$-Higgs theory. The advantage of the lower symmetry model is that it will give us access to more phases, and we can check if our critical theory can be perturbed to obtain the larger phase diagram. The largest symmetry group for 2+1D SMG is $\SO(7)$, where we can still consider a honeycomb lattice model with eight Majorana fermion on each site, transforming like a $\SO(7)$ spinor. The $\SO(7)$ SMG can be driven by applying the  $\SO(7)$ symmetric Fidkowski-Kitaev  interaction\cite{Fidkowski:2010bf,Fidkowski:2011dd} to each site. All the lower symmetry SMGs in 2D can be considered as  descendants of the $\SO(7)$ SMG by partially breaking the $\SO(7)$ symmetry down to its subgroup. An interesting direction is to consider SMG in various dimensions. In 3+1D, once again it is readily shown that eight Dirac fermions (or 16 Weyl fermions) can be gapped to produce a featureless state \cite{Wang:2013my,Wen:2013kr,You:2014ow,You:2015lj,BenTov:2015lh,BenTov:2016co}. Whether this can proceed through a single continuous transition remains to be seen, numerics on one microscopic model appear to give an intervening symmetry breaking phase \cite{Ayyar:2016ph}.  One may also discuss SMG  in  1+1D, where we need four Dirac fermions. In fact this is closely related to the interaction reduction of topological phases in 1+1D described by Fidkowski-Kitaev \cite{Fidkowski:2010bf}, where they show that edge states with eight Majorana modes are unstable despite the presence of time reversal symmetry that forbids a quadratic gapping term. Therefore, the transition between the trivial phase and a phase with eight Majorana (or four Dirac) edge zero modes can be circumvented by interactions, which is related to symmetric mass generation for four Dirac fermions in 1+1D. In Appendix.\,\ref{sec: FK}, we review and reinterpret the Fidkowski-Kitaev transition in 1+1D within an $\SO(7)$ SMG\cite{Fidkowski:2010bf} in the parton language. There, the problem was solved using an alternate set of fermion variables informed by SO(8) triality, within which the transition is simply described. Could such a change of variable or duality transformation be constructed to describe SMG in higher dimensions? These are questions for future work. 

\acknowledgements{We would like to acknowledge the helpful discussions with Max Metlitski, N. Seiberg, T. Senthil, Chong Wang, Andreas Ludwig, John McGreevy, Tarun Grover, Xiao-Liang Qi, Yingfei Gu, Aneesh Manohar, Subir Sachdev and Leon Balents. AV and YZY was supported by a Simons Investigator grant. YCH is supported by a postdoctoral fellowship from the Gordon and Betty Moore Foundation, under the EPiQS initiative, GBMF4306, at Harvard University. CX is supported by the David and Lucile Packard Foundation and NSF Grant No. DMR-1151208.

\appendix

\section{Renormalization Group Analysis}\label{sec: RG}
In this appendix, we present the renormalization group (RG) analysis of the $N_f=2$ $\SU(4)$ QCD theory with short-range fermion interaction. The theory arises from the SMG field theory \eqnref{eq: L} after gapping out the bosonic field $B$. To control the RG calculation, we can generalize the theory to the large-$N_f$ limit. The Lagrangian in consideration reads
\eq{\label{eq: QCD}
\mathcal{L}_f=\sum_{a=1}^{N_f}\sum_{i,j}\sum_{\alpha,\beta}\bar{f}_{a\alpha i}\gamma^\mu_{\alpha\beta}(\ii\partial_\mu-a_\mu^m\tau^m_{ij})f_{a\beta j}+\mathcal{L}_\text{int}.}
The fermionic parton field $f_{a\alpha i}$ is label by the favor index $a=1,\cdots,N_f$, the Dirac index $\alpha=1,2$ and the color index $i=1,2,3,4$. The flavor indices are transformed under the flavor symmetry group $\Sp(N_f/2)$ and the color indices are transformed under the gauge group $\SU(4)$. The case of $N_f=2$ is relevant to our discussion in the main text. $a_\mu^m$ is the $\SU(4)$ gauge field that couples to the fermion via the $\SU(4)$ generators $\tau^m$ acting in the color subspace. The $\SU(2)$ rotation in the Dirac subspace corresponds to the space-time rotation. The $\gamma^\mu$ matrices are chosen as $(\gamma^0,\gamma^1,\gamma^2)=(\sigma^2,\sigma^1,\sigma^3)$ and $\bar{f}_{a\alpha i}=f_{a\beta i}^\dagger \gamma^0_{\beta\alpha}$.

$\mathcal{L}_\text{int}$ denotes the short-range four-fermion interaction. The charge-4e superconducting interaction for the physical fermion will naturally induce a similar interaction for the fermionic parton with the same symmetry properties. It can be verified that the following interaction is the only four-fermion interaction that is invariant under the space-time rotation, the $\Sp(N_f/2)$ symmetry and the $\SU(4)$ gauge transformations, but breaks the $\U(1)$ symmetry in the same manner as the physical fermion interaction.
\eqs{\label{eq: L_int large N}\mathcal{L}_\text{int}&=g(V^{abcd}_{\alpha\beta\gamma\delta}\epsilon_{ijkl}f_{a\alpha i}f_{b\beta j}f_{c\gamma k}f_{d\delta l}+\text{h.c.}),\\
V^{abcd}_{\alpha\beta\gamma\delta}&=J_{ab}J_{cd}\epsilon_{\alpha\beta}\epsilon_{\gamma\delta}+J_{ac}J_{bd}\epsilon_{\alpha\gamma}\epsilon_{\beta\delta}\\
&\phantom{= }+J_{ad}J_{bc}\epsilon_{\alpha\delta}\epsilon_{\beta\gamma},}
where $\epsilon_{ijkl}$ is the totally antisymmetric tensor in the color subspace, $\epsilon_{\alpha\beta}$ is the antisymmetric matrix in the Dirac subspace, and $J_{ab}$ is the symplectic form of the $\Sp(N_f/2)$ group in the flavor subspace (such that the generator $A$ of $\Sp(N_f/2)$ preserves $JA+A^\intercal J=0$).

\begin{figure}[htbp]
\begin{center}
\includegraphics[width=0.4\textwidth]{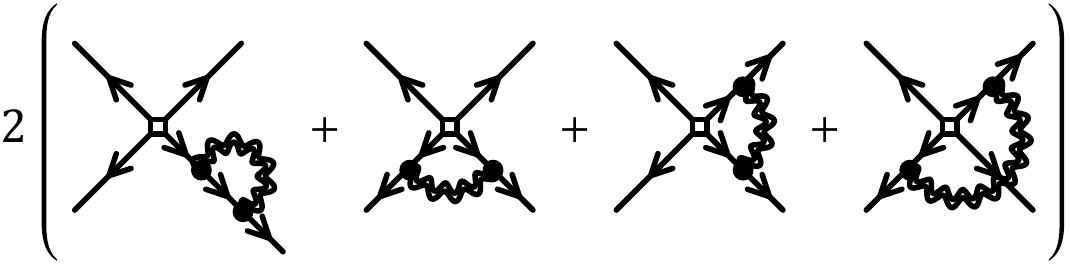}
\caption{Diagrams that contributes the the correction of the interaction vertex. Wavy lines are the gauge boson propagators $D(q)_{\mu\nu}^{mn}=16 q^{-1}(\delta_{\mu\nu}-\xi q_\mu q_\nu/q^2)\delta_{mn}$ at the large-$N_f$ fixed point. The arrowed lines are the fermion propagator $G(k)=1/(k_\mu\gamma^\mu)$.}
\label{fig: RG diagrams}
\end{center}
\end{figure}

\figref{fig: RG diagrams} concludes the diagrams that contributes to the linear order (in $g$) of the RG equation at the $1/N_f$ order. Following Ref.\,\onlinecite{Xu:2008rz,Xu:2008jc}, the RG flow equation is given by
\eq{\frac{\dd g}{\dd \ell}=-\Big(1-\frac{80}{\pi^2N_f}\Big)g.}
So the interaction strength $g$ has the scaling dimension $[g]=-1+80/(\pi^2N_f)+\mathcal{O}(N_f^{-2})$. At $N_f=2$, we have $[g]\approx 3.05>0$, implying that the short-range interaction $g$ is \emph{relevant}. To analyze the  instabilities due to this interaction, we recall the $\vect{\beta}=(\sigma^{12},\sigma^{20},\sigma^{32},\ii\sigma^{21},\ii\sigma^{02},\ii\sigma^{23})$ matrices defined in the main text, then the antisymmetric tensor $\epsilon_{ijkl}$ can be decomposed as $\epsilon_{ijkl}=\frac{1}{2}\sum_{m}\beta_{ij}^m\beta_{kl}^m$. Thus at $N_f=2$, the interaction in \eqnref{eq: L_int large N} becomes
\eq{\mathcal{L}_\text{int}=\frac{g}{2}\big((f_{K}^\intercal\ii\gamma^0\vect{\beta}f_{K'})^2+\text{h.c.}\big).}
So the strong charge-4e interaction could drive the spontaneous generation of the gauge sextet pairing $\vect{\Delta}=f_{K}^\intercal\ii\gamma^0\vect{\beta}f_{K'}$ regardless of the sign of $g$. The sign of $g$ only determines whether the instability is in the $\Re\vect{\Delta}$ channel ($g<0$) or in the $\Im\vect{\Delta}$ channel ($g>0$). But either case will lead to the fermionic parton mass generation and the gauge confinement, so the sign of $g$ is not important.

\section{Derivation of the Green's Function}\label{sec: G(k)}

As we have shown in \eqnref{eq: projective}, in the extreme limit of zero correlation length, the featureless gapped state $\ket{\Psi_c}$ in \eqnref{eq: SC4e} can be obtained exactly by projecting the mean field state $\ket{\Psi_{c,\vect{M}}}$ to the $\SU(4)$ symmetric sector. In this appendix, we would like to generalize this construction to the case of finite correlation length. Although the projective construction will not be exact as we go away from the zero correlation length limit, yet it still provide us an useful variational wave function which has a controlled asymptotically exact limit, based on which we can evaluate the fermion Green's function.

The idea to increase the fermion correlation in the wave function $\ket{\Psi_c}$ is to allow the fermion to move around on the lattice. So we turn on the fermion hopping term in the mean field Hamiltonian,
\eqs{H_\text{MF}&=H_0+H_\vect{M}\\
&= -t\sum_{\langle ij \rangle} c^\dagger_{i}c_{j}-\sum_{i}\vect{M}\cdot\vect{\Delta}_i+\text{h.c.}.}
Let us still denote the mean field ground state as $\ket{\Psi_{c,\vect{M}}}$. Switch to the momentum-frequency space and use the Nambu spinor basis $c_{k}=(c_{K+k},c_{K'+k}^\dagger)^\intercal$, the fermion correlation on the mean field state is given by
\eqs{G_\vect{M}(k)&=-\bra{\Psi_{c,\vect{M}}} c_k \bar{c}_k\ket{\Psi_{c,\vect{M}}}\\
&\simeq \mat{\gamma^\mu k_\mu & -\ii\vect{M}\cdot\vect{\beta}\\
\ii\vect{M}\cdot\vect{\beta} & -\gamma^\mu k_\mu}^{-1}\\
&=\frac{1}{k^\mu k_\mu+\vect{M}^2}\mat{\gamma^\mu k_\mu & -\ii\vect{M}\cdot\vect{\beta}\\
\ii\vect{M}\cdot\vect{\beta} & -\gamma^\mu k_\mu}}
at low energy. We propose an $\SU(4)$ symmetric wave function $\ket{\Psi_c}$ by symmetrizing $\ket{\Psi_{c,\vect{M}}}$ following \eqnref{eq: projective}, as $\ket{\Psi_c}=\int_{S^5} \dd \vect{M}\ket{\Psi_{c,\vect{M}}}$ (assuming the measure is $\SO(6)$ symmetric and is properly normalized), where $S^5$ denotes a sphere of radius $|\vect{M}|$. Then the Green's function on the symmetric state can be obtained by symmetrizing the mean field Green's function. To see this, we start with
\eqs{G(k)&=-\bra{\Psi_c}c_k\bar{c}_k\ket{\Psi_c}\\
&=-\int\dd\vect{M}\dd\vect{M}'\bra{\Psi_{c,\vect{M}}}c_k\bar{c}_k\ket{\Psi_{c,\vect{M}'}}.}
The overlap $\bra{\Psi_{c,\vect{M}}}c_k\bar{c}_k\ket{\Psi_{c,\vect{M}'}}$ vanishes if $\vect{M}\neq\vect{M}'$, due to the orthogonality catastrophe of fermion many-body states. Therefore we have
\eqs{G(k)&=-\int\dd\vect{M}\bra{\Psi_{c,\vect{M}}}c_k\bar{c}_k\ket{\Psi_{c,\vect{M}}}\\
&=\int\dd\vect{M} G_\vect{M}(k).}
The symmetrization will remove the $\vect{M}\cdot\vect{\beta}$ terms in the numerator but leave the $\vect{M}^2$ term in denominator untouched. Switching back from the Nambu basis, we arrived at the Green's function in \eqnref{eq: G(k)}.

\section{Fidkowski-Kitaev SO(7) Symmetric Mass Generation}\label{sec: FK}

In this appendix, we will review the (1+1)D Fidkowski-Kitaev $\SO(7)$ symmetric mass generation\cite{Fidkowski:2010bf,Fidkowski:2011dd} from the perspective of the parton construction. The model is defined on a 1D lattice. On each site, there are eight Majorana fermion modes $\chi_{ia}$ ($a=1,\cdots,8$) forming the 8-dimensional real spinor representation of an $\SO(7)$ group. The Hamiltonian reads $H=H_0+H_I$ with
\eqs{\label{eq: FK}
H_0&=\sum_{i}\sum_{a=1}^{8} \ii\chi_{i,a}\chi_{i+1,a},\\
H_I&=-\frac{V}{4!}\sum_{i}\sum_{m=1}^{7}\Delta_i^m \Delta_i^m,}
where $\Delta_i^m=\chi_{ia}\Gamma^m_{ab}\chi_{ib}$ are the seven components of the $\SO(7)$ vector. The Gamma matrices can be chosen as $\vect{\Gamma}=(\sigma^{123},\sigma^{203},\sigma^{323},\sigma^{211},\sigma^{021},\sigma^{231},\sigma^{002})$, which form a set of purely imaginary, antisymmetric and anticommuting matrices. The Hamiltonian $H$ is manifestly $\SO(7)$ invariant. Besides the internal $\SO(7)$ symmetry, the model also possess the translation symmetry $T:\chi_{i}\to\chi_{i+1}$ and the chiral symmetry $\mathcal{S}:\chi_{i}\to(-)^i\chi_{i},\ii\to-\ii$. One can see $\mathcal{S}^2=+1$ and $T^{-1}\mathcal{S}T\mathcal{S}=-1$ acting on the fermions.

In the non-interacting limit ($V\to0$), $H_0$ simply describes eight decoupled and gapless Majorana chains, whose field theory description is
\eq{\label{eq: Lchi0}\mathcal{L}_0=\frac{1}{2}\sum_{a=1}^{8}\bar{\chi}_a\gamma^\mu\ii\partial_\mu\chi_a,}
where $\chi_a=(\chi_{\text{L}a},\chi_{\text{R}a})^\intercal$ and $\bar{\chi}_a=\chi_a^\intercal \gamma^0$ contains both left- and right-moving Majorana modes with the gamma matrices $(\gamma^0,\gamma^1)=(\ii\sigma^2,\sigma^1)$. Since the Majorana coupling along the chain is purely imaginary, the Fermi points are located at momentum $k=0,\pi$, so $\chi_{\text{L},\text{R}}$ fields are related to the real space fermion $\chi_i$ by
\eq{\chi_{\text{L}a}=\sum_{i}\chi_{ia},\quad \chi_{\text{R}a}=\sum_{i}(-)^i\chi_{ia}.}
Both $\chi_{\text{L}}$ and $\chi_{\text{R}}$ transform as $\SO(7)$ spinors. The translation and the chiral symmetry acts as
\eq{\label{eq: TS sym}T:\chi_{\text{R}a}\to-\chi_{\text{R}a},\quad
\mathcal{S}:\chi_{\text{L}a}\leftrightarrow\chi_{\text{R}a}.}
All fermion bilinear mass terms (such as $\ii m\bar{\chi}\chi$) are forbidden by these symmetries. In fact, the translation symmetry is the most important protecting symmetry which is sufficient to rule out all bilinear masses (i.e.~all back scattering terms between $\chi_\text{L}$ and $\chi_\text{R}$). So the Majorana chain can not be symmetrically gapped on the free fermion level.

However, it is possible to symmetrically gap out eight  Majorana chains by fermion interactions. One possible interaction proposed by Fidkowski and Kitaev\cite{Fidkowski:2010bf,Fidkowski:2011dd} is the $\SO(7)$ symmetric interaction $V$ in \eqnref{eq: FK} (abbreviated as the FK interaction hereinafter). In the strong interaction limit ($V\to+\infty$), the Hamiltonian is dominated by $H_I$, which decouples to each single site. Diagonalizing the on-site Hamiltonian, one finds a unique ground state separated from the excited states by a finite  gap $\Delta=14V$. If we pair up the on-site Majorana fermions into Dirac fermions as $c_{ia}=\chi_{i,2a-1}+\ii\chi_{i,2a}$ ($a=1,\cdots,4$), the ground state wave function can be expressed as
\eq{\ket{\Psi_c}=\prod_{i}(1+c_{i1}^\dagger c_{i2}^\dagger c_{i3}^\dagger c_{i4}^\dagger)\ket{0_c},}
which is the 1D version of the charge-4e superconducting state in \eqnref{eq: SC4e}. It can be verified that $\ket{\Psi_c}$ preserves the full $\SO(7)$, translation and chiral symmetries, and hence a featureless gapped state. The only difference with the (2+1)D case is that in (1+1)D the interaction $V$ is marginal at the free fermion fixed point. Depending on the sign of $V$, it is marginally relevant if $V>0$ and marginally irrelevant if $V<0$. So as long as we turn on an infinitesimal but positive $V$, the system undergoes a 1D version of the symmetric mass generation (SMG) to the featureless gapped phase.

A better understanding of this 1D SMG physics would greatly help us to understand the SMG transitions in all higher dimensions. In the following we will present a parton construction following Ref.\,\onlinecite{Fidkowski:2010bf} using the $\SO(8)$ triality. The $\SO(8)$ triality is a property that the $\SO(8)$ vector $\mathbf{8}$, left-spinor $\mathbf{8}_+$ and right-spinor $\mathbf{8}_-$ representations can fuse to the trivial representation under the trilinear map $t:\mathcal{V}_{\mathbf{8}}\times\mathcal{V}_{\mathbf{8}_+}\times\mathcal{V}_{\mathbf{8}_-}\to\dsR$. The trality map $t$ can also be written as a three-leg tensor $t^m_{ab}$, where the tensor indices $m,a,b$ labels the basis of $\mathbf{8}$, $\mathbf{8}_+$ and $\mathbf{8}_-$ respectively. Without interactions, $H_0$ has the full $\SO(8)$ symmetry that rotates the eight Majorana flavors. The interaction $H_I$ breaks the $\SO(8)$ symmetry down to its $\SO(7)$ subgroup and at the same time fixes $\chi_{a}$ to be one of the spinor representations $\mathbf{8}_\pm$, say $\mathbf{8}_+$. Using the triality tensor $t^m_{ab}$, we can construct the physical fermion $\chi_{a}$ (as $\mathbf{8}_+$ spinor) by fusing the bosonic parton $\phi_{b}$ (as $\mathbf{8}_-$ spinor) and the fermionic parton $\psi_{m}$ (as $\mathbf{8}$ vector) on the field theory level:
\eq{\label{eq: chi}\chi_{Qa}=\sum_{b,m}t^m_{ab}\phi_{Qb}\psi_{Qm},}
which applies to both the left- and right-moving modes $Q=\text{L},\text{R}$. One way to make sense of \eqnref{eq: chi} on the microscopic level is to consider the bosonic parton $\phi_{Qb}$ as a Kondo impurity resting on the boundary of the fermion chain,\cite{Maldacena:1997fk} which is treated as a dynamical variable without spacial dependence. There are two types of Kondo impurities: $\phi_{\text{L}b}$ and $\phi_{\text{R}b}$. The $\phi_{Qb}$ impurity only couples to the $\chi_{Qa}$ fermion and scatters it to the $\psi_{Qm}$ fermion and vice versa. In this fractionalization scheme, both partons carry the $\SO(7)\subset\SO(8)$ symmetry charge, unlike the fractionalization scheme in the main text where only the bosonic parton carries the $\SU(4)$ symmetry charge. The translation and the chiral symmetry acts on the parton fields as follows:
\eq{\label{eq: PSG}
T:\phi_{\text{R}b}\to-\phi_{\text{R}b},\quad
\mathcal{S}: \psi_{\text{L}m}\leftrightarrow\psi_{\text{R}m},}
such that the symmetry action on the physical fermion field in \eqnref{eq: TS sym} can be retrieved. 

Now let us consider putting the fermionic parton $\psi_{m}=(\psi_{\text{L}m},\psi_{\text{R}m})^\intercal$ in the same band structure as the physical fermion $\chi_a=(\chi_{\text{L}a},\chi_{\text{R}a})^\intercal$, described by
\eq{\mathcal{L}_0^\psi=\frac{1}{2}\sum_{m=1}^{8}\bar{\psi}_m\gamma^\mu\ii\partial_\mu\psi_m,}
similar to \eqnref{eq: Lchi0}. As long as the fermionic parton is gapless, the physical fermion is also gapless, which corresponds to the free fermion fixed point. As the FK interaction is turned on, the following interaction for the fermionic parton will be induced:
\eq{\label{eq: psi int}\mathcal{L}_I^\psi=-A\Big(\sum_{m=1}^7\bar{\psi}_m\psi_m\Big)^2-B\Big(\sum_{m=1}^7\bar{\psi}_m\psi_m\Big)\bar{\psi}_8\psi_8,}
which contains two types of short-range interactions ($A$ and $B$ terms). The general form of the parton interaction in \eqnref{eq: psi int} can be argued on symmetry basis.

According to the fractionalization scheme \eqnref{eq: chi}, the fermionic parton $\psi_m$ was assigned to the vector representation of $\SO(8)$. So the $\SO(7)$ symmetry group (as a subgroup of $\SO(8)$) will only rotate seven components of $\psi_m$ and leaving one remaining component invariant. Without loss of generality, we assume $\psi_8$ to be the $\SO(7)$ invariant component, then the $\SO(7)$ vector can be written as ($m=1,\cdots,7$)
\eq{\Delta^m=\ii(\psi_{\text{L}m}\psi_{\text{L}8}+\psi_{\text{R}m}\psi_{\text{R}8}),}
in terms of the fermionic parton bilinear form. As shown in \eqnref{eq: FK}, the FK interaction $H_I\sim - \vect{\Delta}\cdot\vect{\Delta}$ is just a dot product of $\SO(7)$ vectors, so we expect it to induce a same type of interaction for the fermionic partons 
\eq{-B\sum_{m=1}^7\Delta^m\Delta^m=-\frac{B}{2}\Big(\sum_{m=1}^7\bar{\psi}_m\psi_m\Big)\bar{\psi}_8\psi_8.}
This gives rise to the $B$-type of interaction in \eqnref{eq: psi int}. As shown in Ref.\,\onlinecite{Fidkowski:2010bf}, under the RG flow, $A$-type of interaction will be generated with $A>0$ and become relevant. The $A$-type interaction drives a spontaneous mass generation $\langle\sum_{m=1}^7 \ii\bar{\psi}_m\psi_m\rangle=M$ for the first seven $\psi_m$ fermions, which in turn gives rise to the mass $\ii BM\bar{\psi}_8\psi_8$ for the $\psi_8$ fermion via the $B$-type interaction. Hence all the fermionic partons are gapped out via the mass generation. The parton mass $M$ is evidently $\SO(7)$ symmetric. It is also invariant under the translation and the chiral symmetries as seen from \eqnref{eq: PSG}. So in the presence of the FK interaction, the system will enter the featureless gapped phase via spontaneous mass generation for the fermionic partons.

In conclusion, the key point that we learn from this (1+1)D MSG is that the nature of the ``symmetric mass'' for the physical fermion is actually a bilinear mass for the fermionic parton. The fact that the parton bilinear mass does not break the symmetry is either because the fermionic parton is in a different symmetry representation from the physical fermion (e.g.~the $\SO(7)$ symmetry) or because the symmetry charge is carried away by the bosonic parton (e.g.~the translation symmetry). These are the key observations that motivate us to propose the (2+1)D SMG theory in the main text.

\bibliography{SMG}

\begin{thebibliography}{79}
\expandafter\ifx\csname natexlab\endcsname\relax\def\natexlab#1{#1}\fi
\expandafter\ifx\csname bibnamefont\endcsname\relax
  \def\bibnamefont#1{#1}\fi
\expandafter\ifx\csname bibfnamefont\endcsname\relax
  \def\bibfnamefont#1{#1}\fi
\expandafter\ifx\csname citenamefont\endcsname\relax
  \def\citenamefont#1{#1}\fi
\expandafter\ifx\csname url\endcsname\relax
  \def\url#1{\texttt{#1}}\fi
\expandafter\ifx\csname urlprefix\endcsname\relax\def\urlprefix{URL }\fi
\providecommand{\bibinfo}[2]{#2}
\providecommand{\eprint}[2][]{\url{#2}}

\bibitem[{\citenamefont{{Wallace}}(1947)}]{Wallace:1947ei}
\bibinfo{author}{\bibfnamefont{P.~R.} \bibnamefont{{Wallace}}},
  \bibinfo{journal}{Physical Review} \textbf{\bibinfo{volume}{71}},
  \bibinfo{pages}{622} (\bibinfo{year}{1947}).

\bibitem[{\citenamefont{{Divincenzo} and {Mele}}(1984)}]{Divincenzo:1984xu}
\bibinfo{author}{\bibfnamefont{D.~P.} \bibnamefont{{Divincenzo}}}
  \bibnamefont{and} \bibinfo{author}{\bibfnamefont{E.~J.}
  \bibnamefont{{Mele}}}, \bibinfo{journal}{\prb} \textbf{\bibinfo{volume}{29}},
  \bibinfo{pages}{1685} (\bibinfo{year}{1984}).

\bibitem[{\citenamefont{{Murakami}}(2007)}]{Murakami:2007ye}
\bibinfo{author}{\bibfnamefont{S.}~\bibnamefont{{Murakami}}},
  \bibinfo{journal}{New Journal of Physics} \textbf{\bibinfo{volume}{9}},
  \bibinfo{pages}{356} (\bibinfo{year}{2007}), \eprint{0710.0930}.

\bibitem[{\citenamefont{{Wan} et~al.}(2011)\citenamefont{{Wan}, {Turner},
  {Vishwanath}, and {Savrasov}}}]{Wan:2011rp}
\bibinfo{author}{\bibfnamefont{X.}~\bibnamefont{{Wan}}},
  \bibinfo{author}{\bibfnamefont{A.~M.} \bibnamefont{{Turner}}},
  \bibinfo{author}{\bibfnamefont{A.}~\bibnamefont{{Vishwanath}}},
  \bibnamefont{and} \bibinfo{author}{\bibfnamefont{S.~Y.}
  \bibnamefont{{Savrasov}}}, \bibinfo{journal}{\prb}
  \textbf{\bibinfo{volume}{83}}, \bibinfo{eid}{205101} (\bibinfo{year}{2011}),
  \eprint{1007.0016}.

\bibitem[{\citenamefont{{Burkov} and {Balents}}(2011)}]{Burkov:2011ez}
\bibinfo{author}{\bibfnamefont{A.~A.} \bibnamefont{{Burkov}}} \bibnamefont{and}
  \bibinfo{author}{\bibfnamefont{L.}~\bibnamefont{{Balents}}},
  \bibinfo{journal}{Physical Review Letters} \textbf{\bibinfo{volume}{107}},
  \bibinfo{eid}{127205} (\bibinfo{year}{2011}), \eprint{1105.5138}.

\bibitem[{\citenamefont{{Burkov} et~al.}(2011)\citenamefont{{Burkov}, {Hook},
  and {Balents}}}]{Burkov:2011rp}
\bibinfo{author}{\bibfnamefont{A.~A.} \bibnamefont{{Burkov}}},
  \bibinfo{author}{\bibfnamefont{M.~D.} \bibnamefont{{Hook}}},
  \bibnamefont{and}
  \bibinfo{author}{\bibfnamefont{L.}~\bibnamefont{{Balents}}},
  \bibinfo{journal}{\prb} \textbf{\bibinfo{volume}{84}}, \bibinfo{eid}{235126}
  (\bibinfo{year}{2011}), \eprint{1110.1089}.

\bibitem[{\citenamefont{{Young} et~al.}(2012)\citenamefont{{Young}, {Zaheer},
  {Teo}, {Kane}, {Mele}, and {Rappe}}}]{Young:2012pl}
\bibinfo{author}{\bibfnamefont{S.~M.} \bibnamefont{{Young}}},
  \bibinfo{author}{\bibfnamefont{S.}~\bibnamefont{{Zaheer}}},
  \bibinfo{author}{\bibfnamefont{J.~C.~Y.} \bibnamefont{{Teo}}},
  \bibinfo{author}{\bibfnamefont{C.~L.} \bibnamefont{{Kane}}},
  \bibinfo{author}{\bibfnamefont{E.~J.} \bibnamefont{{Mele}}},
  \bibnamefont{and} \bibinfo{author}{\bibfnamefont{A.~M.}
  \bibnamefont{{Rappe}}}, \bibinfo{journal}{Physical Review Letters}
  \textbf{\bibinfo{volume}{108}}, \bibinfo{eid}{140405} (\bibinfo{year}{2012}),
  \eprint{1111.6483}.

\bibitem[{\citenamefont{{Novoselov} et~al.}(2004)\citenamefont{{Novoselov},
  {Geim}, {Morozov}, {Jiang}, {Zhang}, {Dubonos}, {Grigorieva}, and
  {Firsov}}}]{Novoselov:2004hi}
\bibinfo{author}{\bibfnamefont{K.~S.} \bibnamefont{{Novoselov}}},
  \bibinfo{author}{\bibfnamefont{A.~K.} \bibnamefont{{Geim}}},
  \bibinfo{author}{\bibfnamefont{S.~V.} \bibnamefont{{Morozov}}},
  \bibinfo{author}{\bibfnamefont{D.}~\bibnamefont{{Jiang}}},
  \bibinfo{author}{\bibfnamefont{Y.}~\bibnamefont{{Zhang}}},
  \bibinfo{author}{\bibfnamefont{S.~V.} \bibnamefont{{Dubonos}}},
  \bibinfo{author}{\bibfnamefont{I.~V.} \bibnamefont{{Grigorieva}}},
  \bibnamefont{and} \bibinfo{author}{\bibfnamefont{A.~A.}
  \bibnamefont{{Firsov}}}, \bibinfo{journal}{Science}
  \textbf{\bibinfo{volume}{306}}, \bibinfo{pages}{666} (\bibinfo{year}{2004}),
  \eprint{cond-mat/0410550}.

\bibitem[{\citenamefont{{Novoselov} et~al.}(2005)\citenamefont{{Novoselov},
  {Geim}, {Morozov}, {Jiang}, {Katsnelson}, {Grigorieva}, {Dubonos}, and
  {Firsov}}}]{Novoselov:2005ns}
\bibinfo{author}{\bibfnamefont{K.~S.} \bibnamefont{{Novoselov}}},
  \bibinfo{author}{\bibfnamefont{A.~K.} \bibnamefont{{Geim}}},
  \bibinfo{author}{\bibfnamefont{S.~V.} \bibnamefont{{Morozov}}},
  \bibinfo{author}{\bibfnamefont{D.}~\bibnamefont{{Jiang}}},
  \bibinfo{author}{\bibfnamefont{M.~I.} \bibnamefont{{Katsnelson}}},
  \bibinfo{author}{\bibfnamefont{I.~V.} \bibnamefont{{Grigorieva}}},
  \bibinfo{author}{\bibfnamefont{S.~V.} \bibnamefont{{Dubonos}}},
  \bibnamefont{and} \bibinfo{author}{\bibfnamefont{A.~A.}
  \bibnamefont{{Firsov}}}, \bibinfo{journal}{\nat}
  \textbf{\bibinfo{volume}{438}}, \bibinfo{pages}{197} (\bibinfo{year}{2005}),
  \eprint{cond-mat/0509330}.

\bibitem[{\citenamefont{Zhang et~al.}(2005)\citenamefont{Zhang, Tan, Stormer,
  and Kim}}]{Zhang:2005kl}
\bibinfo{author}{\bibfnamefont{Y.}~\bibnamefont{Zhang}},
  \bibinfo{author}{\bibfnamefont{Y.-W.} \bibnamefont{Tan}},
  \bibinfo{author}{\bibfnamefont{H.~L.} \bibnamefont{Stormer}},
  \bibnamefont{and} \bibinfo{author}{\bibfnamefont{P.}~\bibnamefont{Kim}},
  \bibinfo{journal}{Nature} \textbf{\bibinfo{volume}{438}},
  \bibinfo{pages}{201} (\bibinfo{year}{2005}).

\bibitem[{\citenamefont{{Borisenko} et~al.}(2014)\citenamefont{{Borisenko},
  {Gibson}, {Evtushinsky}, {Zabolotnyy}, {B{\"u}chner}, and
  {Cava}}}]{Borisenko:2014yb}
\bibinfo{author}{\bibfnamefont{S.}~\bibnamefont{{Borisenko}}},
  \bibinfo{author}{\bibfnamefont{Q.}~\bibnamefont{{Gibson}}},
  \bibinfo{author}{\bibfnamefont{D.}~\bibnamefont{{Evtushinsky}}},
  \bibinfo{author}{\bibfnamefont{V.}~\bibnamefont{{Zabolotnyy}}},
  \bibinfo{author}{\bibfnamefont{B.}~\bibnamefont{{B{\"u}chner}}},
  \bibnamefont{and} \bibinfo{author}{\bibfnamefont{R.~J.}
  \bibnamefont{{Cava}}}, \bibinfo{journal}{Physical Review Letters}
  \textbf{\bibinfo{volume}{113}}, \bibinfo{eid}{027603} (\bibinfo{year}{2014}),
  \eprint{1309.7978}.

\bibitem[{\citenamefont{{Weng} et~al.}(2015)\citenamefont{{Weng}, {Fang},
  {Fang}, {Bernevig}, and {Dai}}}]{Weng:2015ay}
\bibinfo{author}{\bibfnamefont{H.}~\bibnamefont{{Weng}}},
  \bibinfo{author}{\bibfnamefont{C.}~\bibnamefont{{Fang}}},
  \bibinfo{author}{\bibfnamefont{Z.}~\bibnamefont{{Fang}}},
  \bibinfo{author}{\bibfnamefont{B.~A.} \bibnamefont{{Bernevig}}},
  \bibnamefont{and} \bibinfo{author}{\bibfnamefont{X.}~\bibnamefont{{Dai}}},
  \bibinfo{journal}{Physical Review X} \textbf{\bibinfo{volume}{5}},
  \bibinfo{eid}{011029} (\bibinfo{year}{2015}), \eprint{1501.00060}.

\bibitem[{\citenamefont{{Huang} et~al.}(2015)\citenamefont{{Huang}, {Xu},
  {Belopolski}, {Lee}, {Chang}, {Wang}, {Alidoust}, {Bian}, {Neupane}, {Zhang}
  et~al.}}]{Huang:2015or}
\bibinfo{author}{\bibfnamefont{S.-M.} \bibnamefont{{Huang}}},
  \bibinfo{author}{\bibfnamefont{S.-Y.} \bibnamefont{{Xu}}},
  \bibinfo{author}{\bibfnamefont{I.}~\bibnamefont{{Belopolski}}},
  \bibinfo{author}{\bibfnamefont{C.-C.} \bibnamefont{{Lee}}},
  \bibinfo{author}{\bibfnamefont{G.}~\bibnamefont{{Chang}}},
  \bibinfo{author}{\bibfnamefont{B.}~\bibnamefont{{Wang}}},
  \bibinfo{author}{\bibfnamefont{N.}~\bibnamefont{{Alidoust}}},
  \bibinfo{author}{\bibfnamefont{G.}~\bibnamefont{{Bian}}},
  \bibinfo{author}{\bibfnamefont{M.}~\bibnamefont{{Neupane}}},
  \bibinfo{author}{\bibfnamefont{C.}~\bibnamefont{{Zhang}}},
  \bibnamefont{et~al.}, \bibinfo{journal}{Nature Communications}
  \textbf{\bibinfo{volume}{6}}, \bibinfo{eid}{7373} (\bibinfo{year}{2015}).

\bibitem[{\citenamefont{{Lv} et~al.}(2015)\citenamefont{{Lv}, {Weng}, {Fu},
  {Wang}, {Miao}, {Ma}, {Richard}, {Huang}, {Zhao}, {Chen} et~al.}}]{Lv:2015rv}
\bibinfo{author}{\bibfnamefont{B.~Q.} \bibnamefont{{Lv}}},
  \bibinfo{author}{\bibfnamefont{H.~M.} \bibnamefont{{Weng}}},
  \bibinfo{author}{\bibfnamefont{B.~B.} \bibnamefont{{Fu}}},
  \bibinfo{author}{\bibfnamefont{X.~P.} \bibnamefont{{Wang}}},
  \bibinfo{author}{\bibfnamefont{H.}~\bibnamefont{{Miao}}},
  \bibinfo{author}{\bibfnamefont{J.}~\bibnamefont{{Ma}}},
  \bibinfo{author}{\bibfnamefont{P.}~\bibnamefont{{Richard}}},
  \bibinfo{author}{\bibfnamefont{X.~C.} \bibnamefont{{Huang}}},
  \bibinfo{author}{\bibfnamefont{L.~X.} \bibnamefont{{Zhao}}},
  \bibinfo{author}{\bibfnamefont{G.~F.} \bibnamefont{{Chen}}},
  \bibnamefont{et~al.}, \bibinfo{journal}{Physical Review X}
  \textbf{\bibinfo{volume}{5}}, \bibinfo{eid}{031013} (\bibinfo{year}{2015}),
  \eprint{1502.04684}.

\bibitem[{\citenamefont{{Xu} et~al.}(2015)\citenamefont{{Xu}, {Belopolski},
  {Alidoust}, {Neupane}, {Bian}, {Zhang}, {Sankar}, {Chang}, {Yuan}, {Lee}
  et~al.}}]{Xu:2015eh}
\bibinfo{author}{\bibfnamefont{S.-Y.} \bibnamefont{{Xu}}},
  \bibinfo{author}{\bibfnamefont{I.}~\bibnamefont{{Belopolski}}},
  \bibinfo{author}{\bibfnamefont{N.}~\bibnamefont{{Alidoust}}},
  \bibinfo{author}{\bibfnamefont{M.}~\bibnamefont{{Neupane}}},
  \bibinfo{author}{\bibfnamefont{G.}~\bibnamefont{{Bian}}},
  \bibinfo{author}{\bibfnamefont{C.}~\bibnamefont{{Zhang}}},
  \bibinfo{author}{\bibfnamefont{R.}~\bibnamefont{{Sankar}}},
  \bibinfo{author}{\bibfnamefont{G.}~\bibnamefont{{Chang}}},
  \bibinfo{author}{\bibfnamefont{Z.}~\bibnamefont{{Yuan}}},
  \bibinfo{author}{\bibfnamefont{C.-C.} \bibnamefont{{Lee}}},
  \bibnamefont{et~al.}, \bibinfo{journal}{Science}
  \textbf{\bibinfo{volume}{349}}, \bibinfo{pages}{613} (\bibinfo{year}{2015}),
  \eprint{1502.03807}.

\bibitem[{\citenamefont{{Castro Neto} et~al.}(2009)\citenamefont{{Castro Neto},
  {Guinea}, {Peres}, {Novoselov}, and {Geim}}}]{Castro-Neto:2009dy}
\bibinfo{author}{\bibfnamefont{A.~H.} \bibnamefont{{Castro Neto}}},
  \bibinfo{author}{\bibfnamefont{F.}~\bibnamefont{{Guinea}}},
  \bibinfo{author}{\bibfnamefont{N.~M.~R.} \bibnamefont{{Peres}}},
  \bibinfo{author}{\bibfnamefont{K.~S.} \bibnamefont{{Novoselov}}},
  \bibnamefont{and} \bibinfo{author}{\bibfnamefont{A.~K.}
  \bibnamefont{{Geim}}}, \bibinfo{journal}{Reviews of Modern Physics}
  \textbf{\bibinfo{volume}{81}}, \bibinfo{pages}{109} (\bibinfo{year}{2009}),
  \eprint{0709.1163}.

\bibitem[{\citenamefont{Franz and Molenkamp}(2013)}]{Franz:2013la}
\bibinfo{author}{\bibfnamefont{M.}~\bibnamefont{Franz}} \bibnamefont{and}
  \bibinfo{author}{\bibfnamefont{L.}~\bibnamefont{Molenkamp}},
  \emph{\bibinfo{title}{Topological Insulators}}, vol.~\bibinfo{volume}{6}
  (\bibinfo{publisher}{Elsevier}, \bibinfo{year}{2013}).

\bibitem[{\citenamefont{Gross and Neveu}(1974)}]{Gross:1974fc}
\bibinfo{author}{\bibfnamefont{D.~J.} \bibnamefont{Gross}} \bibnamefont{and}
  \bibinfo{author}{\bibfnamefont{A.}~\bibnamefont{Neveu}},
  \bibinfo{journal}{Phys. Rev. D} \textbf{\bibinfo{volume}{10}},
  \bibinfo{pages}{3235} (\bibinfo{year}{1974}),
  \urlprefix\url{https://link.aps.org/doi/10.1103/PhysRevD.10.3235}.

\bibitem[{\citenamefont{{Fidkowski} and {Kitaev}}(2010)}]{Fidkowski:2010bf}
\bibinfo{author}{\bibfnamefont{L.}~\bibnamefont{{Fidkowski}}} \bibnamefont{and}
  \bibinfo{author}{\bibfnamefont{A.}~\bibnamefont{{Kitaev}}},
  \bibinfo{journal}{\prb} \textbf{\bibinfo{volume}{81}}, \bibinfo{eid}{134509}
  (\bibinfo{year}{2010}), \eprint{0904.2197}.

\bibitem[{\citenamefont{{Fidkowski} and {Kitaev}}(2011)}]{Fidkowski:2011dd}
\bibinfo{author}{\bibfnamefont{L.}~\bibnamefont{{Fidkowski}}} \bibnamefont{and}
  \bibinfo{author}{\bibfnamefont{A.}~\bibnamefont{{Kitaev}}},
  \bibinfo{journal}{\prb} \textbf{\bibinfo{volume}{83}}, \bibinfo{eid}{075103}
  (\bibinfo{year}{2011}), \eprint{1008.4138}.

\bibitem[{\citenamefont{{Ryu} and {Zhang}}(2012)}]{Ryu:2012ph}
\bibinfo{author}{\bibfnamefont{S.}~\bibnamefont{{Ryu}}} \bibnamefont{and}
  \bibinfo{author}{\bibfnamefont{S.-C.} \bibnamefont{{Zhang}}},
  \bibinfo{journal}{\prb} \textbf{\bibinfo{volume}{85}}, \bibinfo{eid}{245132}
  (\bibinfo{year}{2012}), \eprint{1202.4484}.

\bibitem[{\citenamefont{{Qi}}(2013)}]{Qi:2013qe}
\bibinfo{author}{\bibfnamefont{X.-L.} \bibnamefont{{Qi}}},
  \bibinfo{journal}{New Journal of Physics} \textbf{\bibinfo{volume}{15}},
  \bibinfo{eid}{065002} (\bibinfo{year}{2013}), \eprint{1202.3983}.

\bibitem[{\citenamefont{{Yao} and {Ryu}}(2013)}]{Yao:2013yg}
\bibinfo{author}{\bibfnamefont{H.}~\bibnamefont{{Yao}}} \bibnamefont{and}
  \bibinfo{author}{\bibfnamefont{S.}~\bibnamefont{{Ryu}}},
  \bibinfo{journal}{\prb} \textbf{\bibinfo{volume}{88}}, \bibinfo{eid}{064507}
  (\bibinfo{year}{2013}), \eprint{1202.5805}.

\bibitem[{\citenamefont{{Fidkowski} et~al.}(2013)\citenamefont{{Fidkowski},
  {Chen}, and {Vishwanath}}}]{Fidkowski:2013ww}
\bibinfo{author}{\bibfnamefont{L.}~\bibnamefont{{Fidkowski}}},
  \bibinfo{author}{\bibfnamefont{X.}~\bibnamefont{{Chen}}}, \bibnamefont{and}
  \bibinfo{author}{\bibfnamefont{A.}~\bibnamefont{{Vishwanath}}},
  \bibinfo{journal}{Physical Review X} \textbf{\bibinfo{volume}{3}},
  \bibinfo{eid}{041016} (\bibinfo{year}{2013}), \eprint{1305.5851}.

\bibitem[{\citenamefont{{Wang} and {Senthil}}(2014)}]{Wang:2014lm}
\bibinfo{author}{\bibfnamefont{C.}~\bibnamefont{{Wang}}} \bibnamefont{and}
  \bibinfo{author}{\bibfnamefont{T.}~\bibnamefont{{Senthil}}},
  \bibinfo{journal}{\prb} \textbf{\bibinfo{volume}{89}}, \bibinfo{eid}{195124}
  (\bibinfo{year}{2014}), \eprint{1401.1142}.

\bibitem[{\citenamefont{{Gu} and {Levin}}(2014)}]{Gu:2014tw}
\bibinfo{author}{\bibfnamefont{Z.-C.} \bibnamefont{{Gu}}} \bibnamefont{and}
  \bibinfo{author}{\bibfnamefont{M.}~\bibnamefont{{Levin}}},
  \bibinfo{journal}{\prb} \textbf{\bibinfo{volume}{89}}, \bibinfo{eid}{201113}
  (\bibinfo{year}{2014}), \eprint{1304.4569}.

\bibitem[{\citenamefont{{You} and {Xu}}(2014)}]{You:2014vp}
\bibinfo{author}{\bibfnamefont{Y.-Z.} \bibnamefont{{You}}} \bibnamefont{and}
  \bibinfo{author}{\bibfnamefont{C.}~\bibnamefont{{Xu}}},
  \bibinfo{journal}{\prb} \textbf{\bibinfo{volume}{90}}, \bibinfo{eid}{245120}
  (\bibinfo{year}{2014}), \eprint{1409.0168}.

\bibitem[{\citenamefont{{Yoshida} and {Furusaki}}(2015)}]{Yoshida:2015aj}
\bibinfo{author}{\bibfnamefont{T.}~\bibnamefont{{Yoshida}}} \bibnamefont{and}
  \bibinfo{author}{\bibfnamefont{A.}~\bibnamefont{{Furusaki}}},
  \bibinfo{journal}{\prb} \textbf{\bibinfo{volume}{92}}, \bibinfo{eid}{085114}
  (\bibinfo{year}{2015}), \eprint{1505.06598}.

\bibitem[{\citenamefont{{Gu} and {Qi}}(2015)}]{Gu:2015cy}
\bibinfo{author}{\bibfnamefont{Y.}~\bibnamefont{{Gu}}} \bibnamefont{and}
  \bibinfo{author}{\bibfnamefont{X.-L.} \bibnamefont{{Qi}}},
  \bibinfo{journal}{ArXiv e-prints}  (\bibinfo{year}{2015}),
  \eprint{1512.04919}.

\bibitem[{\citenamefont{{Song} and {Schnyder}}(2016)}]{Song:2016ut}
\bibinfo{author}{\bibfnamefont{X.-Y.} \bibnamefont{{Song}}} \bibnamefont{and}
  \bibinfo{author}{\bibfnamefont{A.~P.} \bibnamefont{{Schnyder}}},
  \bibinfo{journal}{ArXiv e-prints}  (\bibinfo{year}{2016}),
  \eprint{1609.07469}.

\bibitem[{\citenamefont{{Queiroz} et~al.}(2016)\citenamefont{{Queiroz},
  {Khalaf}, and {Stern}}}]{Queiroz:2016se}
\bibinfo{author}{\bibfnamefont{R.}~\bibnamefont{{Queiroz}}},
  \bibinfo{author}{\bibfnamefont{E.}~\bibnamefont{{Khalaf}}}, \bibnamefont{and}
  \bibinfo{author}{\bibfnamefont{A.}~\bibnamefont{{Stern}}},
  \bibinfo{journal}{Physical Review Letters} \textbf{\bibinfo{volume}{117}},
  \bibinfo{eid}{206405} (\bibinfo{year}{2016}), \eprint{1601.01596}.

\bibitem[{\citenamefont{{Wang} and {Gu}}(2017)}]{Wang:2017ty}
\bibinfo{author}{\bibfnamefont{Q.-R.} \bibnamefont{{Wang}}} \bibnamefont{and}
  \bibinfo{author}{\bibfnamefont{Z.-C.} \bibnamefont{{Gu}}},
  \bibinfo{journal}{ArXiv e-prints}  (\bibinfo{year}{2017}),
  \eprint{1703.10937}.

\bibitem[{\citenamefont{{Metlitski} et~al.}(2014)\citenamefont{{Metlitski},
  {Fidkowski}, {Chen}, and {Vishwanath}}}]{Metlitski:2014fp}
\bibinfo{author}{\bibfnamefont{M.~A.} \bibnamefont{{Metlitski}}},
  \bibinfo{author}{\bibfnamefont{L.}~\bibnamefont{{Fidkowski}}},
  \bibinfo{author}{\bibfnamefont{X.}~\bibnamefont{{Chen}}}, \bibnamefont{and}
  \bibinfo{author}{\bibfnamefont{A.}~\bibnamefont{{Vishwanath}}},
  \bibinfo{journal}{ArXiv e-prints}  (\bibinfo{year}{2014}),
  \eprint{1406.3032}.

\bibitem[{\citenamefont{{Slagle} et~al.}(2015)\citenamefont{{Slagle}, {You},
  and {Xu}}}]{Slagle:2015lo}
\bibinfo{author}{\bibfnamefont{K.}~\bibnamefont{{Slagle}}},
  \bibinfo{author}{\bibfnamefont{Y.-Z.} \bibnamefont{{You}}}, \bibnamefont{and}
  \bibinfo{author}{\bibfnamefont{C.}~\bibnamefont{{Xu}}},
  \bibinfo{journal}{\prb} \textbf{\bibinfo{volume}{91}}, \bibinfo{eid}{115121}
  (\bibinfo{year}{2015}), \eprint{1409.7401}.

\bibitem[{\citenamefont{{Kimchi} et~al.}(2013)\citenamefont{{Kimchi},
  {Parameswaran}, {Turner}, {Wang}, and {Vishwanath}}}]{Kimchi:2013fk}
\bibinfo{author}{\bibfnamefont{I.}~\bibnamefont{{Kimchi}}},
  \bibinfo{author}{\bibfnamefont{S.~A.} \bibnamefont{{Parameswaran}}},
  \bibinfo{author}{\bibfnamefont{A.~M.} \bibnamefont{{Turner}}},
  \bibinfo{author}{\bibfnamefont{F.}~\bibnamefont{{Wang}}}, \bibnamefont{and}
  \bibinfo{author}{\bibfnamefont{A.}~\bibnamefont{{Vishwanath}}},
  \bibinfo{journal}{Proceedings of the National Academy of Science}
  \textbf{\bibinfo{volume}{110}}, \bibinfo{pages}{16378}
  (\bibinfo{year}{2013}), \eprint{1207.0498}.

\bibitem[{\citenamefont{{Jiang} and {Ran}}(2015)}]{Jiang:2015qv}
\bibinfo{author}{\bibfnamefont{S.}~\bibnamefont{{Jiang}}} \bibnamefont{and}
  \bibinfo{author}{\bibfnamefont{Y.}~\bibnamefont{{Ran}}},
  \bibinfo{journal}{\prb} \textbf{\bibinfo{volume}{92}}, \bibinfo{eid}{104414}
  (\bibinfo{year}{2015}), \eprint{1505.03171}.

\bibitem[{\citenamefont{{Kim} et~al.}(2016)\citenamefont{{Kim}, {Lee}, {Jiang},
  {Ware}, {Jian}, {Zaletel}, {Han}, and {Ran}}}]{Kim:2016mz}
\bibinfo{author}{\bibfnamefont{P.}~\bibnamefont{{Kim}}},
  \bibinfo{author}{\bibfnamefont{H.}~\bibnamefont{{Lee}}},
  \bibinfo{author}{\bibfnamefont{S.}~\bibnamefont{{Jiang}}},
  \bibinfo{author}{\bibfnamefont{B.}~\bibnamefont{{Ware}}},
  \bibinfo{author}{\bibfnamefont{C.-M.} \bibnamefont{{Jian}}},
  \bibinfo{author}{\bibfnamefont{M.}~\bibnamefont{{Zaletel}}},
  \bibinfo{author}{\bibfnamefont{J.~H.} \bibnamefont{{Han}}}, \bibnamefont{and}
  \bibinfo{author}{\bibfnamefont{Y.}~\bibnamefont{{Ran}}},
  \bibinfo{journal}{\prb} \textbf{\bibinfo{volume}{94}}, \bibinfo{eid}{064432}
  (\bibinfo{year}{2016}), \eprint{1509.04358}.

\bibitem[{\citenamefont{{BenTov}}(2015)}]{BenTov:2015lh}
\bibinfo{author}{\bibfnamefont{Y.}~\bibnamefont{{BenTov}}},
  \bibinfo{journal}{Journal of High Energy Physics}
  \textbf{\bibinfo{volume}{7}}, \bibinfo{eid}{34} (\bibinfo{year}{2015}),
  \eprint{1412.0154}.

\bibitem[{\citenamefont{{He} et~al.}(2016)\citenamefont{{He}, {Wu}, {You},
  {Xu}, {Meng}, and {Lu}}}]{He:2016qy}
\bibinfo{author}{\bibfnamefont{Y.-Y.} \bibnamefont{{He}}},
  \bibinfo{author}{\bibfnamefont{H.-Q.} \bibnamefont{{Wu}}},
  \bibinfo{author}{\bibfnamefont{Y.-Z.} \bibnamefont{{You}}},
  \bibinfo{author}{\bibfnamefont{C.}~\bibnamefont{{Xu}}},
  \bibinfo{author}{\bibfnamefont{Z.~Y.} \bibnamefont{{Meng}}},
  \bibnamefont{and} \bibinfo{author}{\bibfnamefont{Z.-Y.} \bibnamefont{{Lu}}},
  \bibinfo{journal}{\prb} \textbf{\bibinfo{volume}{94}}, \bibinfo{eid}{241111}
  (\bibinfo{year}{2016}), \eprint{1603.08376}.

\bibitem[{\citenamefont{{Catterall}}(2016)}]{Catterall:2016sw}
\bibinfo{author}{\bibfnamefont{S.}~\bibnamefont{{Catterall}}},
  \bibinfo{journal}{Journal of High Energy Physics}
  \textbf{\bibinfo{volume}{1}}, \bibinfo{eid}{121} (\bibinfo{year}{2016}),
  \eprint{1510.04153}.

\bibitem[{\citenamefont{{Ayyar} and
  {Chandrasekharan}}(2016{\natexlab{a}})}]{Ayyar:2016fi}
\bibinfo{author}{\bibfnamefont{V.}~\bibnamefont{{Ayyar}}} \bibnamefont{and}
  \bibinfo{author}{\bibfnamefont{S.}~\bibnamefont{{Chandrasekharan}}},
  \bibinfo{journal}{\prd} \textbf{\bibinfo{volume}{93}}, \bibinfo{eid}{081701}
  (\bibinfo{year}{2016}{\natexlab{a}}), \eprint{1511.09071}.

\bibitem[{\citenamefont{{Lee} and {Lee}}(2005)}]{Lee:2005qq}
\bibinfo{author}{\bibfnamefont{S.-S.} \bibnamefont{{Lee}}} \bibnamefont{and}
  \bibinfo{author}{\bibfnamefont{P.~A.} \bibnamefont{{Lee}}},
  \bibinfo{journal}{Physical Review Letters} \textbf{\bibinfo{volume}{95}},
  \bibinfo{eid}{036403} (\bibinfo{year}{2005}), \eprint{cond-mat/0502139}.

\bibitem[{\citenamefont{{Motrunich}}(2005)}]{Motrunich:2005fp}
\bibinfo{author}{\bibfnamefont{O.~I.} \bibnamefont{{Motrunich}}},
  \bibinfo{journal}{\prb} \textbf{\bibinfo{volume}{72}}, \bibinfo{eid}{045105}
  (\bibinfo{year}{2005}), \eprint{cond-mat/0412556}.

\bibitem[{\citenamefont{{Sheng} et~al.}(2009)\citenamefont{{Sheng},
  {Motrunich}, and {Fisher}}}]{Sheng:2009hc}
\bibinfo{author}{\bibfnamefont{D.~N.} \bibnamefont{{Sheng}}},
  \bibinfo{author}{\bibfnamefont{O.~I.} \bibnamefont{{Motrunich}}},
  \bibnamefont{and} \bibinfo{author}{\bibfnamefont{M.~P.~A.}
  \bibnamefont{{Fisher}}}, \bibinfo{journal}{\prb}
  \textbf{\bibinfo{volume}{79}}, \bibinfo{eid}{205112} (\bibinfo{year}{2009}),
  \eprint{0902.4210}.

\bibitem[{\citenamefont{{Block} et~al.}(2011)\citenamefont{{Block}, {Sheng},
  {Motrunich}, and {Fisher}}}]{Block:2011ad}
\bibinfo{author}{\bibfnamefont{M.~S.} \bibnamefont{{Block}}},
  \bibinfo{author}{\bibfnamefont{D.~N.} \bibnamefont{{Sheng}}},
  \bibinfo{author}{\bibfnamefont{O.~I.} \bibnamefont{{Motrunich}}},
  \bibnamefont{and} \bibinfo{author}{\bibfnamefont{M.~P.~A.}
  \bibnamefont{{Fisher}}}, \bibinfo{journal}{Physical Review Letters}
  \textbf{\bibinfo{volume}{106}}, \bibinfo{eid}{157202} (\bibinfo{year}{2011}),
  \eprint{1009.1179}.

\bibitem[{\citenamefont{{Catterall} and {Schaich}}(2016)}]{Catterall:2016nh}
\bibinfo{author}{\bibfnamefont{S.}~\bibnamefont{{Catterall}}} \bibnamefont{and}
  \bibinfo{author}{\bibfnamefont{D.}~\bibnamefont{{Schaich}}},
  \bibinfo{journal}{ArXiv e-prints}  (\bibinfo{year}{2016}),
  \eprint{1609.08541}.

\bibitem[{\citenamefont{{Ayyar} and
  {Chandrasekharan}}(2016{\natexlab{b}})}]{Ayyar:2016tg}
\bibinfo{author}{\bibfnamefont{V.}~\bibnamefont{{Ayyar}}} \bibnamefont{and}
  \bibinfo{author}{\bibfnamefont{S.}~\bibnamefont{{Chandrasekharan}}},
  \bibinfo{journal}{Journal of High Energy Physics}
  \textbf{\bibinfo{volume}{10}}, \bibinfo{eid}{58}
  (\bibinfo{year}{2016}{\natexlab{b}}), \eprint{1606.06312}.

\bibitem[{\citenamefont{{Ayyar}}(2016)}]{Ayyar:2016ph}
\bibinfo{author}{\bibfnamefont{V.}~\bibnamefont{{Ayyar}}},
  \bibinfo{journal}{ArXiv e-prints}  (\bibinfo{year}{2016}),
  \eprint{1611.00280}.

\bibitem[{\citenamefont{{Senthil}
  et~al.}(2004{\natexlab{a}})\citenamefont{{Senthil}, {Vishwanath}, {Balents},
  {Sachdev}, and {Fisher}}}]{Senthil:2004wj}
\bibinfo{author}{\bibfnamefont{T.}~\bibnamefont{{Senthil}}},
  \bibinfo{author}{\bibfnamefont{A.}~\bibnamefont{{Vishwanath}}},
  \bibinfo{author}{\bibfnamefont{L.}~\bibnamefont{{Balents}}},
  \bibinfo{author}{\bibfnamefont{S.}~\bibnamefont{{Sachdev}}},
  \bibnamefont{and} \bibinfo{author}{\bibfnamefont{M.~P.~A.}
  \bibnamefont{{Fisher}}}, \bibinfo{journal}{Science}
  \textbf{\bibinfo{volume}{303}}, \bibinfo{pages}{1490}
  (\bibinfo{year}{2004}{\natexlab{a}}), \eprint{cond-mat/0311326}.

\bibitem[{\citenamefont{{Motrunich} and {Vishwanath}}(2004)}]{Motrunich:2004hh}
\bibinfo{author}{\bibfnamefont{O.~I.} \bibnamefont{{Motrunich}}}
  \bibnamefont{and}
  \bibinfo{author}{\bibfnamefont{A.}~\bibnamefont{{Vishwanath}}},
  \bibinfo{journal}{\prb} \textbf{\bibinfo{volume}{70}}, \bibinfo{eid}{075104}
  (\bibinfo{year}{2004}), \eprint{cond-mat/0311222}.

\bibitem[{\citenamefont{{Senthil}
  et~al.}(2004{\natexlab{b}})\citenamefont{{Senthil}, {Balents}, {Sachdev},
  {Vishwanath}, and {Fisher}}}]{Senthil:2004qm}
\bibinfo{author}{\bibfnamefont{T.}~\bibnamefont{{Senthil}}},
  \bibinfo{author}{\bibfnamefont{L.}~\bibnamefont{{Balents}}},
  \bibinfo{author}{\bibfnamefont{S.}~\bibnamefont{{Sachdev}}},
  \bibinfo{author}{\bibfnamefont{A.}~\bibnamefont{{Vishwanath}}},
  \bibnamefont{and} \bibinfo{author}{\bibfnamefont{M.~P.~A.}
  \bibnamefont{{Fisher}}}, \bibinfo{journal}{\prb}
  \textbf{\bibinfo{volume}{70}}, \bibinfo{eid}{144407}
  (\bibinfo{year}{2004}{\natexlab{b}}), \eprint{cond-mat/0312617}.

\bibitem[{\citenamefont{{Kivelson} et~al.}(1990)\citenamefont{{Kivelson},
  {Emery}, and {Lin}}}]{Kivelson:1990wy}
\bibinfo{author}{\bibfnamefont{S.~A.} \bibnamefont{{Kivelson}}},
  \bibinfo{author}{\bibfnamefont{V.~J.} \bibnamefont{{Emery}}},
  \bibnamefont{and} \bibinfo{author}{\bibfnamefont{H.~Q.} \bibnamefont{{Lin}}},
  \bibinfo{journal}{\prb} \textbf{\bibinfo{volume}{42}}, \bibinfo{pages}{6523}
  (\bibinfo{year}{1990}).

\bibitem[{\citenamefont{{Berg} et~al.}(2009{\natexlab{a}})\citenamefont{{Berg},
  {Fradkin}, and {Kivelson}}}]{Berg:2009uq}
\bibinfo{author}{\bibfnamefont{E.}~\bibnamefont{{Berg}}},
  \bibinfo{author}{\bibfnamefont{E.}~\bibnamefont{{Fradkin}}},
  \bibnamefont{and} \bibinfo{author}{\bibfnamefont{S.~A.}
  \bibnamefont{{Kivelson}}}, \bibinfo{journal}{\prb}
  \textbf{\bibinfo{volume}{79}}, \bibinfo{eid}{064515}
  (\bibinfo{year}{2009}{\natexlab{a}}), \eprint{0810.1564}.

\bibitem[{\citenamefont{{Radzihovsky} and
  {Vishwanath}}(2009)}]{Radzihovsky:2009bk}
\bibinfo{author}{\bibfnamefont{L.}~\bibnamefont{{Radzihovsky}}}
  \bibnamefont{and}
  \bibinfo{author}{\bibfnamefont{A.}~\bibnamefont{{Vishwanath}}},
  \bibinfo{journal}{Physical Review Letters} \textbf{\bibinfo{volume}{103}},
  \bibinfo{eid}{010404} (\bibinfo{year}{2009}), \eprint{0812.3945}.

\bibitem[{\citenamefont{{Berg} et~al.}(2009{\natexlab{b}})\citenamefont{{Berg},
  {Fradkin}, and {Kivelson}}}]{Berg:2009nm}
\bibinfo{author}{\bibfnamefont{E.}~\bibnamefont{{Berg}}},
  \bibinfo{author}{\bibfnamefont{E.}~\bibnamefont{{Fradkin}}},
  \bibnamefont{and} \bibinfo{author}{\bibfnamefont{S.~A.}
  \bibnamefont{{Kivelson}}}, \bibinfo{journal}{Nature Physics}
  \textbf{\bibinfo{volume}{5}}, \bibinfo{pages}{830}
  (\bibinfo{year}{2009}{\natexlab{b}}), \eprint{0904.1230}.

\bibitem[{\citenamefont{{Moon}}(2012)}]{Moon:2012ng}
\bibinfo{author}{\bibfnamefont{E.-G.} \bibnamefont{{Moon}}},
  \bibinfo{journal}{\prb} \textbf{\bibinfo{volume}{85}}, \bibinfo{eid}{245123}
  (\bibinfo{year}{2012}), \eprint{1202.5389}.

\bibitem[{\citenamefont{{Jiang} et~al.}(2016)\citenamefont{{Jiang}, {Li},
  {Kivelson}, and {Yao}}}]{Jiang:2016km}
\bibinfo{author}{\bibfnamefont{Y.-F.} \bibnamefont{{Jiang}}},
  \bibinfo{author}{\bibfnamefont{Z.-X.} \bibnamefont{{Li}}},
  \bibinfo{author}{\bibfnamefont{S.~A.} \bibnamefont{{Kivelson}}},
  \bibnamefont{and} \bibinfo{author}{\bibfnamefont{H.}~\bibnamefont{{Yao}}},
  \bibinfo{journal}{ArXiv e-prints}  (\bibinfo{year}{2016}),
  \eprint{1607.01770}.

\bibitem[{\citenamefont{{You} and {Xu}}(2015)}]{You:2015lj}
\bibinfo{author}{\bibfnamefont{Y.-Z.} \bibnamefont{{You}}} \bibnamefont{and}
  \bibinfo{author}{\bibfnamefont{C.}~\bibnamefont{{Xu}}},
  \bibinfo{journal}{\prb} \textbf{\bibinfo{volume}{91}}, \bibinfo{eid}{125147}
  (\bibinfo{year}{2015}), \eprint{1412.4784}.

\bibitem[{\citenamefont{{Schnyder} et~al.}(2008)\citenamefont{{Schnyder},
  {Ryu}, {Furusaki}, and {Ludwig}}}]{Schnyder:2008os}
\bibinfo{author}{\bibfnamefont{A.~P.} \bibnamefont{{Schnyder}}},
  \bibinfo{author}{\bibfnamefont{S.}~\bibnamefont{{Ryu}}},
  \bibinfo{author}{\bibfnamefont{A.}~\bibnamefont{{Furusaki}}},
  \bibnamefont{and} \bibinfo{author}{\bibfnamefont{A.~W.~W.}
  \bibnamefont{{Ludwig}}}, \bibinfo{journal}{\prb}
  \textbf{\bibinfo{volume}{78}}, \bibinfo{eid}{195125} (\bibinfo{year}{2008}),
  \eprint{0803.2786}.

\bibitem[{\citenamefont{{Ryu} et~al.}(2010)\citenamefont{{Ryu}, {Schnyder},
  {Furusaki}, and {Ludwig}}}]{Ryu:2010fe}
\bibinfo{author}{\bibfnamefont{S.}~\bibnamefont{{Ryu}}},
  \bibinfo{author}{\bibfnamefont{A.~P.} \bibnamefont{{Schnyder}}},
  \bibinfo{author}{\bibfnamefont{A.}~\bibnamefont{{Furusaki}}},
  \bibnamefont{and} \bibinfo{author}{\bibfnamefont{A.~W.~W.}
  \bibnamefont{{Ludwig}}}, \bibinfo{journal}{New Journal of Physics}
  \textbf{\bibinfo{volume}{12}}, \bibinfo{eid}{065010} (\bibinfo{year}{2010}),
  \eprint{0912.2157}.

\bibitem[{\citenamefont{{Ludwig}}(2016)}]{Ludwig:2016pt}
\bibinfo{author}{\bibfnamefont{A.~W.~W.} \bibnamefont{{Ludwig}}},
  \bibinfo{journal}{Physica Scripta Volume T} \textbf{\bibinfo{volume}{168}},
  \bibinfo{eid}{014001} (\bibinfo{year}{2016}), \eprint{1512.08882}.

\bibitem[{\citenamefont{{Wu}}(2006)}]{Wu:2006gf}
\bibinfo{author}{\bibfnamefont{C.}~\bibnamefont{{Wu}}},
  \bibinfo{journal}{Modern Physics Letters B} \textbf{\bibinfo{volume}{20}},
  \bibinfo{pages}{1707} (\bibinfo{year}{2006}), \eprint{cond-mat/0608690}.

\bibitem[{\citenamefont{{Grover}}(2012)}]{Grover:2012rw}
\bibinfo{author}{\bibfnamefont{T.}~\bibnamefont{{Grover}}},
  \bibinfo{journal}{ArXiv e-prints}  (\bibinfo{year}{2012}),
  \eprint{1211.1392}.

\bibitem[{\citenamefont{{Vafa} and {Witten}}(1984)}]{VafaWitten}
\bibinfo{author}{\bibfnamefont{C.}~\bibnamefont{{Vafa}}} \bibnamefont{and}
  \bibinfo{author}{\bibfnamefont{E.}~\bibnamefont{{Witten}}},
  \bibinfo{journal}{Nuclear Physics B} \textbf{\bibinfo{volume}{234}},
  \bibinfo{pages}{173} (\bibinfo{year}{1984}).

\bibitem[{\citenamefont{{Cai} et~al.}(2013{\natexlab{a}})\citenamefont{{Cai},
  {Hung}, {Wang}, {Zheng}, and {Wu}}}]{Cai:2013ak}
\bibinfo{author}{\bibfnamefont{Z.}~\bibnamefont{{Cai}}},
  \bibinfo{author}{\bibfnamefont{H.-h.} \bibnamefont{{Hung}}},
  \bibinfo{author}{\bibfnamefont{L.}~\bibnamefont{{Wang}}},
  \bibinfo{author}{\bibfnamefont{D.}~\bibnamefont{{Zheng}}}, \bibnamefont{and}
  \bibinfo{author}{\bibfnamefont{C.}~\bibnamefont{{Wu}}},
  \bibinfo{journal}{Physical Review Letters} \textbf{\bibinfo{volume}{110}},
  \bibinfo{eid}{220401} (\bibinfo{year}{2013}{\natexlab{a}}),
  \eprint{1202.6323}.

\bibitem[{\citenamefont{{Cai} et~al.}(2013{\natexlab{b}})\citenamefont{{Cai},
  {Hung}, {Wang}, and {Wu}}}]{Cai:2013mq}
\bibinfo{author}{\bibfnamefont{Z.}~\bibnamefont{{Cai}}},
  \bibinfo{author}{\bibfnamefont{H.-H.} \bibnamefont{{Hung}}},
  \bibinfo{author}{\bibfnamefont{L.}~\bibnamefont{{Wang}}}, \bibnamefont{and}
  \bibinfo{author}{\bibfnamefont{C.}~\bibnamefont{{Wu}}},
  \bibinfo{journal}{\prb} \textbf{\bibinfo{volume}{88}}, \bibinfo{eid}{125108}
  (\bibinfo{year}{2013}{\natexlab{b}}), \eprint{1207.6843}.

\bibitem[{\citenamefont{{Wang} et~al.}(2014)\citenamefont{{Wang}, {Li}, {Cai},
  {Zhou}, {Wang}, and {Wu}}}]{Wang:2014kl}
\bibinfo{author}{\bibfnamefont{D.}~\bibnamefont{{Wang}}},
  \bibinfo{author}{\bibfnamefont{Y.}~\bibnamefont{{Li}}},
  \bibinfo{author}{\bibfnamefont{Z.}~\bibnamefont{{Cai}}},
  \bibinfo{author}{\bibfnamefont{Z.}~\bibnamefont{{Zhou}}},
  \bibinfo{author}{\bibfnamefont{Y.}~\bibnamefont{{Wang}}}, \bibnamefont{and}
  \bibinfo{author}{\bibfnamefont{C.}~\bibnamefont{{Wu}}},
  \bibinfo{journal}{Physical Review Letters} \textbf{\bibinfo{volume}{112}},
  \bibinfo{eid}{156403} (\bibinfo{year}{2014}), \eprint{1305.3571}.

\bibitem[{\citenamefont{{You} et~al.}(2014{\natexlab{a}})\citenamefont{{You},
  {Wang}, {Oon}, and {Xu}}}]{You:2014br}
\bibinfo{author}{\bibfnamefont{Y.-Z.} \bibnamefont{{You}}},
  \bibinfo{author}{\bibfnamefont{Z.}~\bibnamefont{{Wang}}},
  \bibinfo{author}{\bibfnamefont{J.}~\bibnamefont{{Oon}}}, \bibnamefont{and}
  \bibinfo{author}{\bibfnamefont{C.}~\bibnamefont{{Xu}}},
  \bibinfo{journal}{\prb} \textbf{\bibinfo{volume}{90}}, \bibinfo{eid}{060502}
  (\bibinfo{year}{2014}{\natexlab{a}}), \eprint{1403.4938}.

\bibitem[{\citenamefont{{Gurarie}}(2011)}]{Gurarie:2011gl}
\bibinfo{author}{\bibfnamefont{V.}~\bibnamefont{{Gurarie}}},
  \bibinfo{journal}{\prb} \textbf{\bibinfo{volume}{83}}, \bibinfo{eid}{085426}
  (\bibinfo{year}{2011}), \eprint{1011.2273}.

\bibitem[{\citenamefont{{Senthil} and {Fisher}}(2006)}]{Senthil:2006cs}
\bibinfo{author}{\bibfnamefont{T.}~\bibnamefont{{Senthil}}} \bibnamefont{and}
  \bibinfo{author}{\bibfnamefont{M.~P.~A.} \bibnamefont{{Fisher}}},
  \bibinfo{journal}{\prb} \textbf{\bibinfo{volume}{74}}, \bibinfo{eid}{064405}
  (\bibinfo{year}{2006}), \eprint{cond-mat/0510459}.

\bibitem[{\citenamefont{{Wang} et~al.}(2017)\citenamefont{{Wang}, {Nahum},
  {Metlitski}, {Xu}, and {Senthil}}}]{Wang:2017zp}
\bibinfo{author}{\bibfnamefont{C.}~\bibnamefont{{Wang}}},
  \bibinfo{author}{\bibfnamefont{A.}~\bibnamefont{{Nahum}}},
  \bibinfo{author}{\bibfnamefont{M.~A.} \bibnamefont{{Metlitski}}},
  \bibinfo{author}{\bibfnamefont{C.}~\bibnamefont{{Xu}}}, \bibnamefont{and}
  \bibinfo{author}{\bibfnamefont{T.}~\bibnamefont{{Senthil}}},
  \bibinfo{journal}{ArXiv e-prints}  (\bibinfo{year}{2017}),
  \eprint{1703.02426}.

\bibitem[{\citenamefont{{You} et~al.}(2017)\citenamefont{{You}, {He},
  {Vishwanath}, and {Xu}}}]{You:2017sy}
\bibinfo{author}{\bibfnamefont{Y.-Z.} \bibnamefont{{You}}},
  \bibinfo{author}{\bibfnamefont{Y.-C.} \bibnamefont{{He}}},
  \bibinfo{author}{\bibfnamefont{A.}~\bibnamefont{{Vishwanath}}},
  \bibnamefont{and} \bibinfo{author}{\bibfnamefont{C.}~\bibnamefont{{Xu}}},
  \bibinfo{journal}{ArXiv e-prints}  (\bibinfo{year}{2017}),
  \eprint{1711.00863}.

\bibitem[{\citenamefont{{Wang} and {Wen}}(2013)}]{Wang:2013my}
\bibinfo{author}{\bibfnamefont{J.}~\bibnamefont{{Wang}}} \bibnamefont{and}
  \bibinfo{author}{\bibfnamefont{X.-G.} \bibnamefont{{Wen}}},
  \bibinfo{journal}{ArXiv e-prints}  (\bibinfo{year}{2013}),
  \eprint{1307.7480}.

\bibitem[{\citenamefont{{Wen}}(2013)}]{Wen:2013kr}
\bibinfo{author}{\bibfnamefont{X.-G.} \bibnamefont{{Wen}}},
  \bibinfo{journal}{Chinese Physics Letters} \textbf{\bibinfo{volume}{30}},
  \bibinfo{eid}{111101} (\bibinfo{year}{2013}), \eprint{1305.1045}.

\bibitem[{\citenamefont{{You} et~al.}(2014{\natexlab{b}})\citenamefont{{You},
  {BenTov}, and {Xu}}}]{You:2014ow}
\bibinfo{author}{\bibfnamefont{Y.-Z.} \bibnamefont{{You}}},
  \bibinfo{author}{\bibfnamefont{Y.}~\bibnamefont{{BenTov}}}, \bibnamefont{and}
  \bibinfo{author}{\bibfnamefont{C.}~\bibnamefont{{Xu}}},
  \bibinfo{journal}{ArXiv e-prints}  (\bibinfo{year}{2014}{\natexlab{b}}),
  \eprint{1402.4151}.

\bibitem[{\citenamefont{{BenTov} and {Zee}}(2016)}]{BenTov:2016co}
\bibinfo{author}{\bibfnamefont{Y.}~\bibnamefont{{BenTov}}} \bibnamefont{and}
  \bibinfo{author}{\bibfnamefont{A.}~\bibnamefont{{Zee}}},
  \bibinfo{journal}{\prd} \textbf{\bibinfo{volume}{93}}, \bibinfo{eid}{065036}
  (\bibinfo{year}{2016}), \eprint{1505.04312}.

\bibitem[{\citenamefont{{Xu} and {Sachdev}}(2008)}]{Xu:2008rz}
\bibinfo{author}{\bibfnamefont{C.}~\bibnamefont{{Xu}}} \bibnamefont{and}
  \bibinfo{author}{\bibfnamefont{S.}~\bibnamefont{{Sachdev}}},
  \bibinfo{journal}{Physical Review Letters} \textbf{\bibinfo{volume}{100}},
  \bibinfo{eid}{137201} (\bibinfo{year}{2008}), \eprint{0711.3460}.

\bibitem[{\citenamefont{{Xu}}(2008)}]{Xu:2008jc}
\bibinfo{author}{\bibfnamefont{C.}~\bibnamefont{{Xu}}}, \bibinfo{journal}{\prb}
  \textbf{\bibinfo{volume}{78}}, \bibinfo{eid}{054432} (\bibinfo{year}{2008}),
  \eprint{0803.0794}.

\bibitem[{\citenamefont{{Maldacena} and {Ludwig}}(1997)}]{Maldacena:1997fk}
\bibinfo{author}{\bibfnamefont{J.~M.} \bibnamefont{{Maldacena}}}
  \bibnamefont{and} \bibinfo{author}{\bibfnamefont{A.~W.~W.}
  \bibnamefont{{Ludwig}}}, \bibinfo{journal}{Nuclear Physics B}
  \textbf{\bibinfo{volume}{506}}, \bibinfo{pages}{565} (\bibinfo{year}{1997}),
  \eprint{cond-mat/9502109}.

\end{thebibliography}
\bibliographystyle{apsrev}
\end{document}